\documentclass[twocolumn]{aastex63}
\usepackage{natbib}
\usepackage{amssymb}
\usepackage{amsmath}
\usepackage{verbatim}
\usepackage{indentfirst}
\usepackage{xcolor}
\usepackage{graphicx}

\usepackage{lineno}

\newcommand{\mbh}{M$_{\rm BH}$}
\newcommand{\Ha}{H$\alpha$}
\newcommand{\msun}{M$_{\odot}$} 
\newcommand{\kms}{km s$^{-1}$} 

\shortauthors{Shin et al.}

\newcommand{\RNum}[1]{\uppercase\expandafter{\romannumeral #1\relax}}
\usepackage{amsmath}

\newcommand{\snu}{\affil{Department of Physics \& Astronomy, Seoul National University, Seoul 08826, Republic of Korea}}
\newcommand{\snuarc}{\affil{SNU Astronomy Research Center, Seoul National University, Seoul 08826, Republic of Korea}}
\newcommand{\nysc}{\affil{National Youth Space Center, Goheung 59567, Republic of Korea}}
\newcommand{\UM}{\affil{Department of Astronomy, University of Michigan, 1085 S University Ave, Ann Arbor, MI 48109, USA}}
\begin{document}

\title{Search of Intermediate Mass Black Holes at Low Redshift with Intra-night Variability}
\correspondingauthor{Jong-Hak Woo}\email{woo@astro.snu.ac.kr}

\author{Lael~Shin} \snu
\author[0000-0002-8055-5465]{Jong-Hak~Woo} \snu\snuarc
\author{Donghoon~Son} \snu
\author[0000-0003-2010-8521]{Hojin~Cho} \snu
\author{Taewoo~Kim} \nysc
\author{Elena Gallo} \UM
\author{Wonseok~Kang} \nysc

\begin{abstract}
We present a sample of intermediate-mass black hole (IMBH) candidates based on the detection of a broad \Ha\ emission line and variability, which are selected from the Sloan Digital Sky Survey Data Release 7. By performing spectral decomposition of emission lines as well as visual inspection, we initially identified 131 targets with a broad \Ha\ line among a large sample of emission-line galaxies. We further selected 25 IMBH candidates, whose estimated black hole mass (\mbh) is less than $10^6 \rm M_{\odot}$. 
To constrain the nature of these candidates, we analyzed X-ray properties and performed an intra-night variability monitoring with optical telescopes.
Based on the optical variability analysis, we report a sample of 11 targets with detected intra-night variability as the best IMBH candidates, which are suitable for follow-up observations for accurate \mbh\ determination such as reverberation mapping campaigns.

\end{abstract}

\keywords{galaxies: active --- galaxies: nuclei --- techniques: photometric}

\section{Introduction\label{sec:intro}}
The decades-long observations have established that supermassive black holes (SMBHs) with mass $>$10$^6$ \msun\ are universal in the center of massive galaxies at low redshift (e.g., see \citealt{Kormendy&Richstone95}; \citealt{Kormendy&Ho13}), yet the origin and the mass range of the high-redshift seeds of these SMBHs are unclear. The existence of stellar mass black holes (BHs) in a much lower mass range $\sim$10-100 $\rm M_{\odot}$ has been well studied by various observations of electromagnetic radiation as well as gravitational wave \citep[e.g.,][]{Remillard&McClintock06, Thompson+19, Abbott+20}. However, discovery of $\sim10^9 \rm M_{\odot}$ SMBH in the very early universe triggered
a fundamental question on the BH formation and seed mass (\citealt{Mortlock+11}; \citealt{Banados+18}; \citealt{Onoue+19}). One of the scenarios suggests that the progenitor of the high-z SMBHs is much heavier than normal stellar mass BH \citep[e.g.,][]{Loeb&Rasio94, Latif+13, Greene+20}.

Thus, by bridging the two well-observed \mbh\ ranges, systematic search of intermediate-mass black holes (IMBHs) with masses of $10^3-10^6\rm M_{\odot}$ in the present-day universe is pivotal to understand how the seeds of SMBH formed and evolved throughout cosmic time  \citep[e.g.,][]{Volonteri+08, Greene12, Greene+20}. If some of the seed BHs at high redshift remain as IMBHs without significantly growing over cosmic time, the relics of the seed can be observed in nearby galaxies as IMBHs \citep[e.g.,][]{Greene12, Reines&Comastri16, Mezcua17, Woo+19a}. 

Nevertheless, IMBHs are rarely observed, hence, they are called the ‘missing link’ between stellar mass BHs and SMBHs (see review by \citealt{Mezcua17}). The evidence of IMBH candidates has been found from various observations. First, X-ray emission and the dynamical modeling of a number of globular clusters (GCs) suggested the presence of an IMBH in their center(e.g., \citealt{Gebhardt+00}; \citealt{Gerssen+02}; \citealt{Pooley&Rappaport06}). For example, \citet{Silk&Arons75} first argued that the X-ray flux from their GC targets could be explained by the accretion of the IMBHs with masses in the range 100-1000$ \rm M_{\odot}$. \citet{Baumgardt17} also claimed that IMBH with mass of  $\sim4\times 10^4 \rm M_{\odot}$ exists in $\omega$ Centauri by analyzing the velocity dispersion profile of the stars in the GC. Tens of IMBH candidates were identified in similar ways, but their methodologies depended on the complicated models of kinematics and radiation, limited to nearby IMBH candidates. Second, extragalactic ultra-luminous X-ray sources (ULXs) were proposed as IMBHs. Black hole mass estimation based on the fundamental plane of accreting black holes (e.g.,  \citealt{Webb+12}; \citealt{Cseh+15}) and the X-ray spectrum fitting (e.g., \citealt{Miller+03}; \citealt{Sutton+12}) indicated that several BHs were in the IMBH mass range. These methods required multi-wavelength observations, including X-ray and radio, along with complex modeling of the accretion disk. 

Detection of the broad emission lines, which are one of the main signatures of mass-accreting type 1 active galactic nuclei (AGN), also provided evidence for IMBH candidates \citep[e.g.,][]{Filippenko&Ho03,Greene&Ho04,Greene&Ho07, Dong+12}. Compared with the other methods, this method enables discovery of IMBH candidates in distant galaxies, and the mass of these candidates can be better determined. Assuming that the gas in the broad line region (BLR) is virialized and governed by the gravitational potential of the central black hole, the mass of the black holes can be determined based on the virial relation:
\begin{equation}
\label{eq:virial}
M_{BH} = f \times R_{BLR}\Delta V^2/G
\end{equation}
where $R_{BLR}$ is the BLR size, $\Delta V$ is the velocity of gas measured from the width of broad emission lines, and $f$ is a dimensionless scale factor depending on the kinematics and geometry of the BLR \citep[e.g.,][]{Peterson93, Peterson+04}. Adopting the correlation between AGN continuum luminosity (or emission line luminosity) and the BLR size \citep[e.g.,][]{Bentz+13}, the black hole mass can be indirectly estimated based on single-epoch spectra \citep[e.g.,][]{Woo&Urry02, Greene&ho05, Park+12, Shen&liu12, Le+20}.

Among the broad emission lines, the H$\alpha$ line is often utilized to identify type 1 AGNs because of its high intensity \citep[e.g.,][]{Woo+14, Eun+17}. While typical type 1 AGNs with strong \Ha\ have \mbh\ larger than 10$^6$ \msun, a relatively weak broad H$\alpha$ line should be present in AGNs with intermediate-mass. Thus, a number of studies utilized the presence of a weak broad \Ha\ to search IMBH \citep[e.g.,][]{Greene&Ho04, Greene&Ho07, Dong+12, Reines+13, Liu+18, Chilingarian+18}.

For example, \cite{Reines+13} identified IMBH candidates by detecting a weak broad component of the H$\alpha$ emission line,
using a sample of galaxies at z $<$ 0.055, whose stellar mass is smaller than $3\times10^9 \rm M_{\odot}$ by assuming that IMBHs are more likely to exist in dwarf galaxies as suggested by black hole mass - stellar mass relation \citep[e.g.,][]{Greene12, McConnell&Ma13, Reines&Comastri16}. Based on the broad \Ha, they reported 17 IMBH candidates, however, additional studies are required to confirm their mass because the mass estimation based on the single-epoch spectrum analysis suffers from large uncertainty, and it is possible that other sources such as luminous blue variable stars may be responsible for the broad \Ha\ \citep[e.g.,][]{Izotov&Thuan07, Izotov&Thuan09a}. 

The true nature of IMBH candidates can be constrained by other properties of type 1 AGNs. For example, detection of the high-luminosity of X-ray emission can provide a constraint to confirm whether they are mass-accreting BHs since AGNs are sources of strong X-ray radiation (e.g., \citealt{Elvis+78}; \citealt{Nucita+17}; \citealt{Chilingarian+18}; \citealt{Liu+18}). Variability also provides strong constraints as the AGN continuum and emission line flux changes over time (e.g., \citealt{Rodriguez-Pascual+97}; \citealt{Walsh+09}; \citealt{Rakshit&Stalin17}; \citealt{Baldassare+18}; \citealt{Martinez+20}). In the case of IMBH, variability test over intra-night or several day time scale is a crucial tool for identification of these candidates (e.g., \citealt{Mushotzky+11}; \citealt{Aranzana+18}; \citealt{Kim+18}).

In this paper we present the results of our systematic search of IMBH candidates. By detecting the H$\alpha$ emission line and focusing on the targets with the weakest H$\alpha$ broad component, we identified a sample of IMBH candidates. Then, we investigated X-ray properties and intra-night variability. We describe sample selection in Section \ref{sec:sample}, observations and the data reduction in Section \ref{sec:observation}, and variability analysis in Section \ref{sec:variability}. We provide discussion in Section \ref{sec:dis}, and summary and conclusion are followed in Section \ref{sec:con}.

%%% FIGURE %%%%%%%%%%%%%%%%%%%%%%%%%%%%%%%%%%%%%%%%%%%%%%%%%%%%%%%%%%%%%%%%%%%%
\begin{figure*}[t]
\centering
\includegraphics[angle=0,width=175mm]{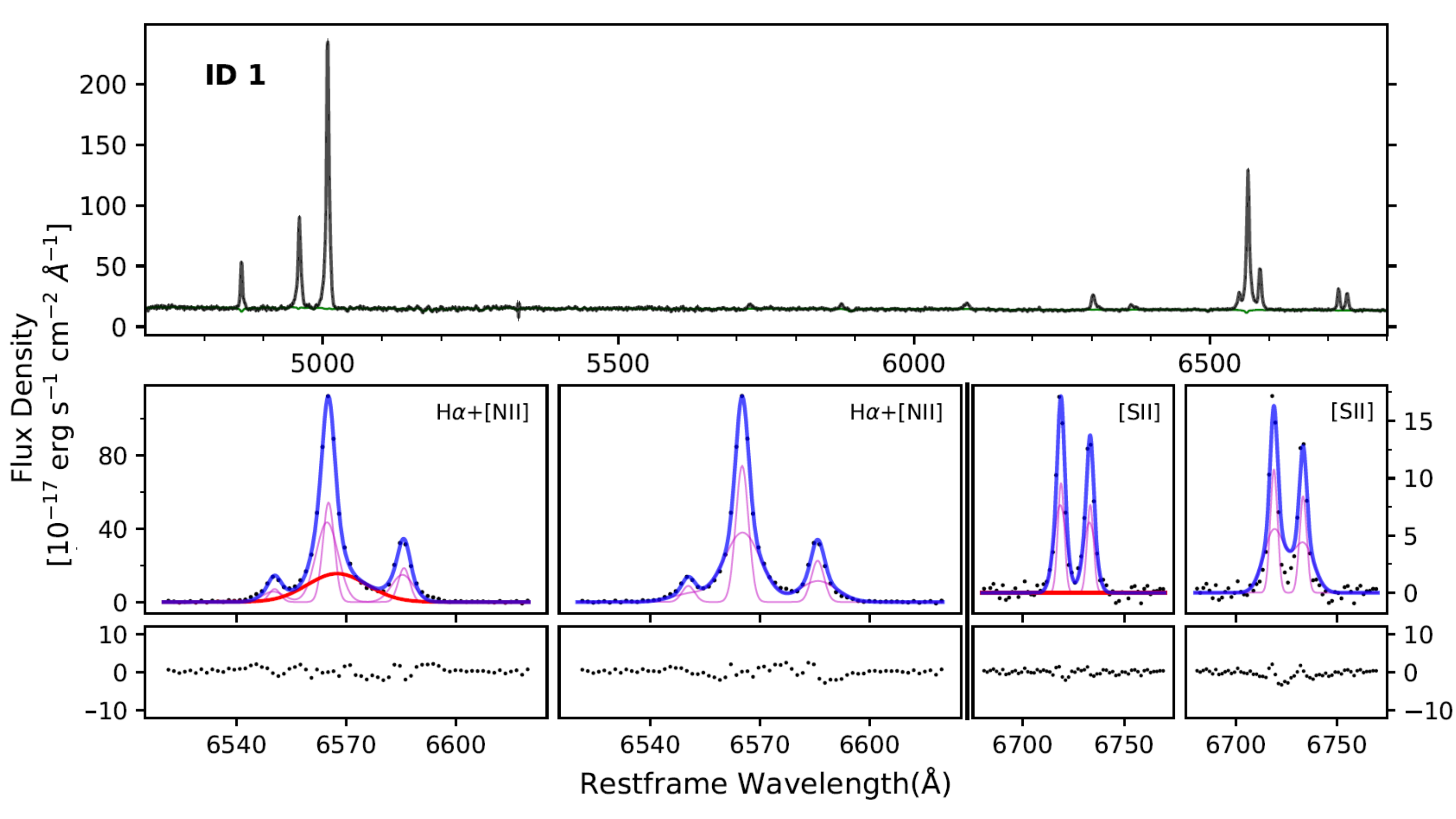}
\caption{Example of a broad H$\alpha$ candidate (ID 1) whose estimated M$_{\rm BH}$ is smaller than $10^{6} \rm M_{\odot}$. The top panel shows the redshift-corrected spectrum with the best stellar continuum fit plotted in green. In the bottom panels, the first and third column show the best fit model with H$\alpha$ broad component for each emission line region. The second and forth column show the best fit model without H$\alpha$ broad component. The stellar continuum subtracted spectrum is plotted black, and the blue line shows the best fit model composed of narrow Gaussian components (magenta) and H$\alpha$ broad component (red). At the bottom of each panel, the residuals between the spectrum and the best fit model are plotted in black. Spectra of the other 24 broad H$\alpha$ candidates with estimated M$_{\rm BH}$ $<$ $10^{6}$ $ \rm M_{\odot}$ are shown in the Appendix. \label{fig:fitting_25}}
\end{figure*}
%%%%%%%%%%%%%%%%%%%%%%%%%%%%%%%%%%%%%%%%%%%%%%%%%%%%%%%%%%%%%%%%%%%%%%%%%%%%%%%

\begin{deluxetable*}{llllllll}[!ht]

\tablecaption{X-ray detected broad H$\alpha$ candidates \label{tab:table_xray}}
\tablewidth{0.95\textwidth}
\tablecolumns{8}
\tablehead
{
\colhead{SDSS Name} & \colhead{Mission} & \colhead{E range} & \colhead{$\log\;F_{X-ray}$} & \colhead{Distance} & \colhead{$\log\;L_{X-ray}$} & \colhead{$\log\;L_{H\alpha}$} & \colhead{$\log\;M_{BH}$} \\
\colhead{(1)} & \colhead{(2)} & \colhead{(3)} & \colhead{(4)} & \colhead{(5)} & \colhead{(6)} & \colhead{(7)} & \colhead{(8)}
}

\startdata
J075953.48+232324.2 &  XMM-Newton &   0.2-12 &    -11.0 &      132 &       43.3 &     40.2 &  6.1 \\
J083736.97+245959.2 &  XMM-Newton &   0.2-12 &    -13.2 &      131 &       41.1 &     39.9 &  6.1 \\
J084344.98+354942.0 &       Swift &   15-150 &    -11.0 &      241 &       43.9 &     41.0 &  6.2 \\
J090229.38+032305.9 &     Chandra &  0.3-8.0 &    -12.8 &      124 &       41.5 &     39.9 &  6.2 \\
J094319.14+361452.1 &       ROSAT &  0.1-2.4 &    -12.5 &      102 &       41.6 &     40.2 &  6.4 \\
J110306.23+554100.0 &       ROSAT &  0.1-2.4 &    -12.4 &      217 &       42.3 &     40.2 &  6.5 \\
J110501.97+594103.6 &     Chandra &  0.3-8.0 &    -11.9 &      151 &       42.6 &     40.8 &  6.7 \\
J112301.31+470308.6 &  XMM-Newton &   0.2-12 &    -14.0 &      114 &       40.2 &     40.0 &  6.1 \\
J114612.17+202329.9 &     Chandra &  0.3-8.0 &    -11.9 &      108 &       42.2 &     40.4 &  6.4 \\
J133514.41+104110.2 &  XMM-Newton &   0.2-12 &    -11.7 &      180 &       42.9 &     40.4 &  6.7 \\
J134632.13+642325.1 &  XMM-Newton &   0.2-12 &    -11.0 &      108 &       43.2 &     40.3 &  6.3 \\
J135419.95+325547.7$^a$ &  XMM-Newton &   0.2-12 &    -11.2 &      118 &       43.0 &     40.7 &  6.8 \\
&     Chandra &  0.3-8.0 &    -12.0 &          &       42.2 &          &      \\
J141451.34+030751.3 &    Einstein &   0.2-4  &    -12.6 &      115 &       41.6 &     40.0 &  6.2 \\
J143318.47+344404.4 &     Chandra &  0.3-8.0 &    -12.5 &      154 &       42.0 &     40.2 &  6.5 \\
J143642.70+341837.5 &     Chandra &  0.3-8.0 &    -13.4 &      197 &       41.2 &     39.9 &  6.2 \\
J144108.69+351958.8 &        ASCA &    0.7-7 &    -12.4 &      352 &       42.8 &     40.9 &  6.3 \\
J144958.69+522801.3 &       ROSAT &  0.1-2.4 &    -13.1 &      421 &       42.2 &     40.9 &  6.4 \\
J151405.52+094209.7 &       ROSAT &  0.1-2.4 &    -12.8 &      347 &       42.3 &     40.4 &  6.6 \\
J212512.47-071329.8 &     Chandra &  0.3-8.0 &    -12.3 &      280 &       42.6 &     40.6 &  6.6 \\
\enddata
\tablecomments{(1) SDSS name. (2) X-ray mission name. (3) Energy range of the X-ray observation in units of keV. (4) Absorption-corrected X-ray flux in units of erg s$^{-1}$ cm$^{-2}$. (5) Distance derived from the redshift in units of Mpc. (6) Absorption-corrected X-ray luminosity in units of erg s$^{-1}$. (7) Luminosity of the H$\alpha$ broad component estimated in this study in units of erg s$^{-1}$. (8) Black hole mass estimated in this study.\\
$^a$Observed in both missions XMM-Newton and Chandra.
}
\end{deluxetable*}

%%%%%%%%%%%%%%%%%%%%%%%%%%%%%%%%%%%%%%%%%%%%%%%%%%%%%%%%%%%%%%%%%%%%%%%%%%%%%%%
\section{Sample Selection\label{sec:sample}}

\subsection{Local AGN $\&$ Star-forming Galaxy Sample\label{sec:ini_sample}}
To find IMBH candidates by detecting a weak H$\alpha$ broad component, we started with a large sample of emission line galaxies from \citet{Bae&Woo14} and \citet{Woo+16}, who classified 60,018 type2 AGN-host galaxies and 128,951 star-forming galaxies (SFGs) based on the flux ratios of the emission lines \citep{Kauffmann+03}, using the Max Planck institute for astrophysics and the Johns Hopkins university (MPA-JHU) catalog of SDSS DR7 galaxies \citep{Abazajian+09}.
For robust spectral analysis, they restricted the sample with two criteria: (1) signal-to-noise ratio (S/N) $>$ 3 for the four emission lines, H$\beta$, [O\RNum{3}] $\lambda$5007, H$\alpha$, and [N\RNum{2}] $\lambda$6584; (2) amplitude(peak)-to-noise ratio (A/N) $>$ 5 for both H$\alpha$, and [O\RNum{3}] $\lambda$5007 emission lines. Based on these criteria, we selected 10,958 type 2 AGNs at z$<$0.1. We also obtained 22,000 SFGs at z $<$ 0.1 by limiting stellar mass of the host galaxy M$_{*}$ $<$ $10^{10}$ \msun, in order to survey dwarf galaxies, which are more likely to host IMBHs  \citep[e.g.,][]{McConnell&Ma13, Reines&Comastri16}.

\subsection{IMBH candidates with a Broad H$\alpha$ Component\label{sec:bro_sample}}
\subsubsection{Spectral decomposition\label{sec:spec}}

We performed spectral decomposition to investigate the presence of a broad \Ha\ component using the selected AGNs and SFGs. It is crucial to decompose AGN emission lines from the host galaxy stellar continuum for properly measuring the flux and velocity dispersion of \Ha. We first subtracted the stellar continuum from the SDSS spectra by using the penalized pixel-fitting (pPXF) routine \citep{Cappellari&Emsellem04}, which finds the best-fit stellar template for the given galaxy spectra. For this process, we used MILES simple stellar population models with solar metallicity \citep{Sanchez-Blazquez+06}. Before the fitting process, we masked the optical emission lines and the H$\alpha$ emission line region (i.e., 6300-6900$\textrm{\AA}$) to prevent fitting a potential broad emission line as a stellar continuum. We then subtracted the best-fit continuum model from the observed spectra, leaving the pure emission line spectra.

Then, we constructed the emission line model for the region around H$\alpha$ (i.e., 6400-6800$\textrm{\AA}$), where [N\RNum{2}] $\lambda$6548, [N\RNum{2}] $\lambda$6584, [S\RNum{2}] $\lambda$6717, and [S\RNum{2}] $\lambda$6731 lines are also located.
To confirm the presence of a broad H$\alpha$ component, we used four different fitting schemes: (a) which fit the narrow component of each emission line ([N\RNum{2}], [S\RNum{2}], and \Ha) with a single Gaussian model, without including an additional Gaussian model for a potential broad H$\alpha$ line, (b) which fit the narrow component with a double Gaussian model without including an additional broad H$\alpha$ component, (c) which fit the narrow component with a single Gaussian model and add one more Gaussian model for a broad H$\alpha$ line, (d) which fit the narrow component with a double Gaussian model and include an additional model for a broad H$\alpha$ line. 
We fixed the relative separations between the centers of each emission line model to the laboratory value, and the width of each narrow emission line to the equal value assuming that they are emitted from the cloud with the same kinematical properties. Furthermore, in the double Gaussian model for narrow components, we added a constraint that the amplitude ratio of the two Gaussian components should be equal for each emission line model.

For modeling, we used \texttt{curve$\_$fit} tool of python \texttt{scipy} package \citep{Virtanen+20}, which utilizes trust region reflective algorithm. To evaluate the reliability of the fitting results, we calculated their reduced chi square ($\chi^2_{red}$) and Bayesian information criterion (BIC) as
\begin{equation}
\label{eq:redchi}
\chi^2_{red} = \frac{1}{n-m}\sum_{i}\frac{\left(O_i - C_i\right)^2}{\sigma_i^2}
\end{equation}
\begin{equation}
\label{eq:bic}
BIC = \sum_{i}\frac{\left(O_i - C_i\right)^2}{\sigma_i^2} + m\ln n
\end{equation}
where $O_i$ is the observed flux density at each wavelength, $C_i$ is the model flux density, and $\sigma_i$ refers to the flux density error. They both consider the number of the data (n) and the number of the model parameters (m). Similar to $\chi^2_{red}$, a better fitting model has less BIC. However, BIC has a penalty term ($m\ln n$) for the number of the model parameters, so a more reliable comparison can be done between the models with different number of parameters. When the difference between the BIC values of two models is greater than 10, it is rated as a strong evidence against the significance of the model with higher BIC \citep{Kass&Raftery95}. 

Based on the results, we set the criteria to select the candidates with a broad H$\alpha$ line. To prevent the overfitting problem, we employed single Gaussian model for the narrow component if the BIC of model (a) is less than the BIC of model (b) + 10, if not, used double Gaussian model for the narrow component. We initially selected the candidates which have a broad H$\alpha$ line using criteria as follows.
\begin{itemize}
\item velocity dispersion of H$\alpha$ broad component after correcting for the SDSS instrumental resolution (60 km s$^{-1}$) $>$ 250 km s$^{-1}$ (i.e. $\sim$590 km s$^{-1}$ FWHM)
\item A/N of H$\alpha$ broad component $>$ 20 (10 for SFGs)
\item BIC of the model with H$\alpha$ broad component + 10 $<$  BIC of the model without H$\alpha$ broad component
\item $\chi^2_{red}$ of the model with H$\alpha$ broad component + 10 $<$  $\chi^2_{red}$ of the model without H$\alpha$ broad component
\end{itemize}
The adopted lower limit of the velocity width (i.e., velocity dispersion 250 \kms\ or FWHM = $\sim$590 \kms) seems lower than the typical definition of the type 1 AGNs (i.e., $>$1000 km s$^{-1}$ FWHM; e.g., \cite{VandenBerk+06}). However, velocity dispersion of 250 km s$^{-1}$ is much larger than stellar velocity dispersion due to gravitational potential of low-mass galaxies. Instead, the line broadening is likely caused by the gravitation of the central black hole. Thus, we interpreted that the line component broader than 250 km s$^{-1}$ is originated from the BLR of the central black hole. There is little concern that the narrow and broad component are misclassified since the narrow line of all candidate is much narrower than 250 km s$^{-1}$. Note that our approach of the velocity lower limit is similar to that of the previous study by \citet{Reines+13}, who surveyed IMBH candidates with a lower limit of FWHM = 500 km s$^{-1}$ (i.e., velocity dispersion of $\sim$210 km s$^{-1}$). 
In addition, we applied smaller amplitude-to-noise ratio cut (A/N $>$ 10) for SFGs than the A/N of AGNs to include more SFGs. Based on these criteria, we obtained 1,484 type 2 AGNs and 116 SFGs for further analysis. 

We visually inspected the spectral decomposition results and the presence of H$\alpha$ broad component for these $\sim$1,500 targets. First, we checked the decomposition result without including a broad H$\alpha$ model. If there is no significant residual around the \Ha\ and [S\RNum{2}] region, we assume that no broad \Ha\ component is present, excluding the target from the candidate list, since it is not clear whether a H$\alpha$ broad component is required in the best-fit model. We present an example of emission line decomposition results with or without a broad H$\alpha$ model in Figure \ref{fig:fitting_25}. Because the width of each emission line model is held fixed by the same value, the model without a broad H$\alpha$ component makes the [S\RNum{2}] line model much broader than the best-model.

There are also examples of the opposite case shown in the Appendix. If a model without a broad H$\alpha$ model fits the [S\RNum{2}] lines, it makes the H$\alpha$ line model much narrower than the best-model, so a significant residual is present around the H$\alpha$ line. In contrast, the model with a broad H$\alpha$ component fits both [S\RNum{2}] and H$\alpha$ emission line better.
Based on the visual inspection, we finalized 738 candidates from type 2 AGNs and 4 candidates from SFGs. 

\subsubsection{Black hole mass estimation}
We estimated black hole mass (M$_{\rm BH}$) of the sample of 738 type 1 AGNs with a broad \Ha,
to select the IMBH candidates. Note that 611 targets among 738 objects were previously studied by \citet{Eun+17}. Since their estimated black hole mass is larger than $10^6 \rm M_{\odot}$, we excluded them from the sample of IMBH candidates. 
For remaining 131 AGNs, we determine M$_{\rm BH}$ using a single-epoch estimator based on the virial assumption and the BLR size - luminosity ($L_{5100}$) relation \citep{Bentz+13}. Because the H$\alpha$ luminosity can be used as a proxy of the AGN continuum luminosity \citep{Greene&ho05}:
\begin{equation}
\label{eq:l5100}
\begin{split}
L_{H\alpha} = &(5.25 \pm 0.02) \\
&\times 10^{42} \left(\frac{L_{5100}}{10^{44}  \:\mathrm{erg}\:\mathrm{s}^{-1}}\right)^{(1.157 \pm 0.005)}\:\mathrm{erg} \:\mathrm{s}^{-1},
\end{split}
\end{equation}
we determine M$_{\rm BH}$ using the velocity dispersion ($\sigma_{H\alpha}$) and luminosity ($L_{H\alpha}$) of the H$\alpha$ broad component \citep{Woo+15}:
\begin{equation}
\label{eq:mbh}
\begin{split}
M_{BH} = f \times 10^{(6.56 \pm 0.06)}&\left(\frac{L_{H\alpha}}{10^{42}\:\mathrm{erg} \:\mathrm{s}^{-1}}\right)^{(0.46 \pm 0.03)} \\
\times &\left(\frac{\sigma_{H\alpha}}{10^{3}\:\mathrm{km} \:\mathrm{s}^{-1}}\right)^{(2.06 \pm 0.06)} M_\odot,
\end{split}
\end{equation}
where $\log f = 0.65 \pm 0.12$ is the virial factor.

We determined the error of each model parameter by calculating the square root of the diagonal term of the covariance matrix provided by \texttt{curve\_fit} tool, and calculated the measurement uncertainty of the M$_{\rm BH}$ using standard propagation of the errors. Considering the intrinsic scatter of the size-luminosity relation \citep[0.19 dex;][]{Bentz+13} and the uncertainty of the virial factor \citep[0.12 dex;][]{Woo+15} as  systematic uncertainties, we obtained the total error of  M$_{\rm BH}$ by adding the measurement and systematic uncertainties in quadrature. The combined error is typically $\sim$0.24 dex.

\begin{figure}
\centering
\includegraphics[width=80mm]{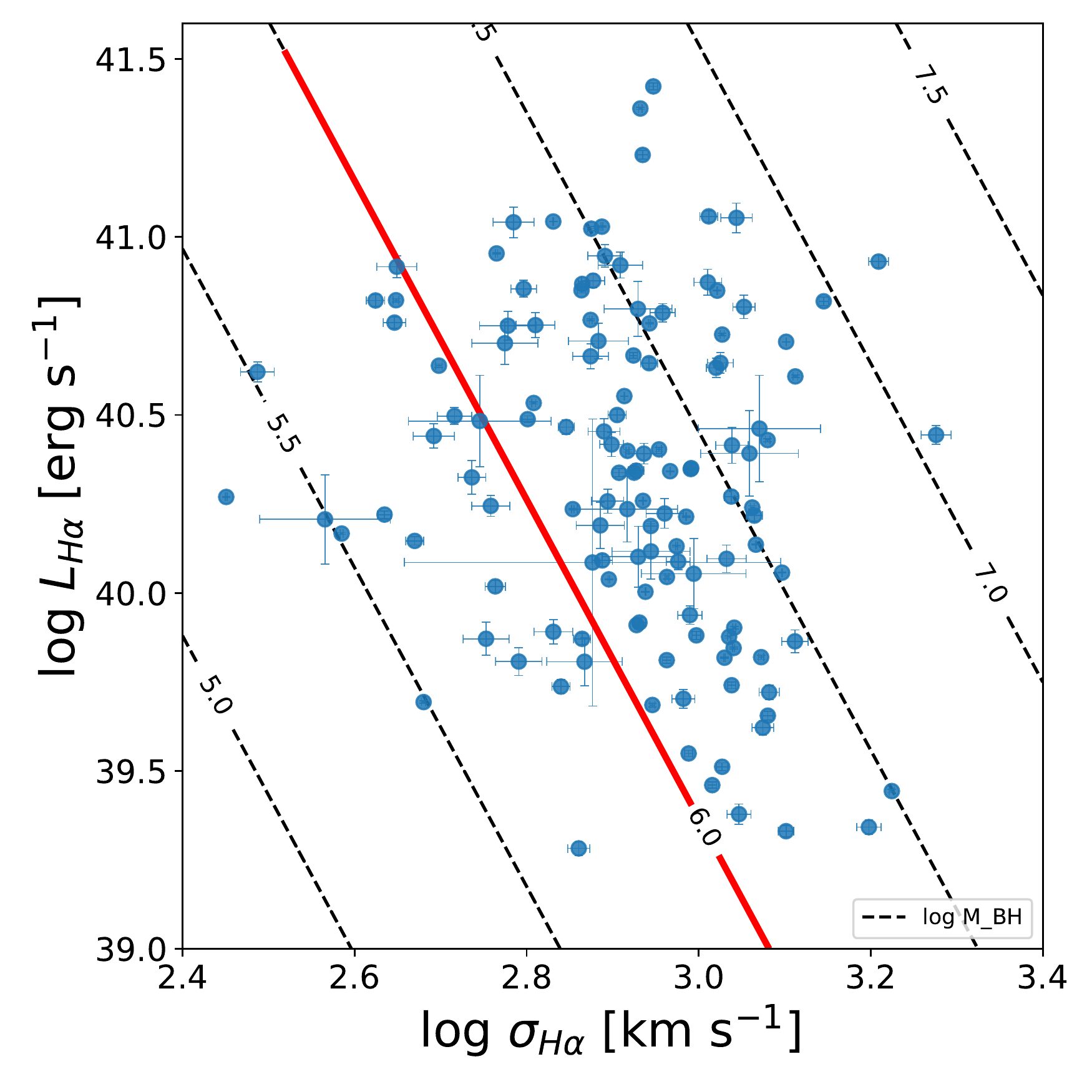}
\caption{Luminosity and velocity dispersion of the H$\alpha$ broad component for the 131 broad H$\alpha$ candidates. Slanted black lines are equi-M$_{\rm BH}$ lines. Red line is the selection criteria, $10^6 \rm M_{\odot}$, for the IMBH candidates. \label{fig:sigl}}
\end{figure}

In Figure \ref{fig:sigl}, we present the distribution of the H$\alpha$ luminosity, velocity dispersion and M$_{\rm BH}$ of the 131 type 1 AGNs. The estimated \mbh\ is in the range $\sim$10$^{5.3}-10^{7.2}\:$ \msun\ with a median of $10^{6.3}\:$ \msun, while we found no target with M$_{\rm BH}$ $<$ $10^{5}$ \msun.
Among them, we identified 25 targets with \mbh\ $<$ 10$^{6}$ \msun\ as IMBH candidates.

\subsection{X-ray black hole candidates}

To find additional evidences that the selected 131 targets with a broad \Ha\ line are AGNs, we investigated X-ray properties by retrieving the X-ray flux data from the online archive NASA/IPAC Extragalactic Database\footnote{https://ned.ipac.caltech.edu/} and NASA HEASARC Xamin Web interface\footnote{https://heasarc.gsfc.nasa.gov/xamin/}. For 19 targerts we obtained X-ray flux data between 0.1 and 150 keV using archival data of various X-ray missions (see Table \ref{tab:table_xray}). The measured X-ray luminosity ranges from $\sim10^{40}$  to $10^{44}$ erg s$^{-1}$ with a median luminosity of $\sim10^{42.3}$ erg s$^{-1}$. However, the estimated M$_{\rm BH}$ of these targets are larger than $10^{6}$ \msun, resulting in no IMBH candidate with available X-ray luminosity. Note that the targets with M$_{\rm BH}$ close to $10^6$ \msun\ can be IMBHs if we consider relatively large uncertainty of the estimated M$_{\rm BH}$. 

\subsection{Pilot sample for a variability test\label{sec:pilot_sample}}

%%% FIGURE %%%%%%%%%%%%%%%%%%%%%%%%%%%%%%%%%%%%%%%%%%%%%%%%%%%%%%%%%%%%%%%%%%%%
%%%%%%%%%%%%%%%%%%%%%%%%%%%%%%%%%%%%%%%%%%%%%%%%%%%%%%%%%%%%%%%%%%%%%%%%%%%%%%%

Using the selected 25 IMBH candidates, we performed optical variability test in order to confirm them as AGNs. In addition, we also observed the IMBH candidates from \citet{Reines+13} as complementary candidates, for which no optical variability test has been performed by \cite{Reines+13}. By applying the same criteria (M$_{\rm BH}$ $<$ $10^{6}$ \msun) to the sample of \citet{Reines+13}, we found 16 targets as a IMBH candidate. Only one target is overlapped with our sample of 25 objects. Note that there are difference between the work of \citet{Reines+13} and ours due to the details of the spectral decomposition. Unlike this study, for example, they subtracted the stellar continuum using different stellar model templates from \citet{Tremonti+04}, and fitted [S\RNum{2}] and H$\alpha$+[N\RNum{2}] composite separately. In addition, they only used the reduced $\chi^2$ to evaluate the spectral decomposition results. The fitting range, fixed parameters for the emission line model, stellar mass limit (i.e., $<3\times10^9 \rm M_{\odot}$) and the redshift range (z $<$ 0.055) are also different. Excluding this one overlapped target, we finalized a sample of 40 IMBH candidates for the variability test observations.

To setup monitoring strategy, we estimated the expected time lag between AGN continuum and H$\alpha$ emission line ($\tau$) based on the BLR size - luminosity relation \citep{Bentz+13}, after adopting the H$\alpha$ luminosity as a proxy of continuum luminosity (Equation \ref{eq:l5100}). The estimated time lag ranges from 0.3 to 5.3 days with a median of 1.9 days. We assign a priority to the candidates with a shorter time lag since they are more likely to have lower \mbh.

\begin{longrotatetable}
\begin{deluxetable*}{llllllllllll}%[!hptb]
\tablecaption{Observation information of the observed sample \label{tab:table_obs}}
\tablecolumns{12}
\tablehead
{
\colhead{ID} & \colhead{NSAID} & \colhead{RA} & \colhead{Dec} & \colhead{z} & \colhead{Filter\_cont} & \colhead{Filter\_$\mathrm{H}\alpha$} & \colhead{Date} & \colhead{Exp\_time} & \colhead{\#\:epoch} & \colhead{Observatory} & \colhead{Reference} \\
&&&&&&&\colhead{(yy.mm.dd)} & \colhead{(cont/$\mathrm{H}\alpha$)} & \colhead{(cont./$\mathrm{H}\alpha$)}&&\\
\colhead{(1)} & \colhead{(2)} & \colhead{(3)} & \colhead{(4)} & \colhead{(5)} & \colhead{(6)} & \colhead{(7)} & \colhead{(8)} & \colhead{(9)} & \colhead{(10)} & \colhead{(11)} & \colhead{(12)}
}
\startdata
1 &   10779 &  09:06:13.8 &  +56:10:15.2 &  0.047 &   kp1494 &  kp1517 &   20.01.08 &           600/600 &               4/4 &         MDM &     This study \\
3 &   33430 &  08:07:07.2 &  +36:14:00.5 &  0.032 &   kp1494 &  kp1497 &   20.01.04 &           600/600 &               5/5 &         MDM &     This study \\
4 &   17134 &  09:55:40.5 &  +05:02:36.7 &  0.034 &   kp1494 &  kp1497 &   20.01.08 &           600/600 &               3/3 &         MDM &     This study \\
5$^a$ &   33232 &  13:08:41.7 &  +52:46:27.4 &  0.024 &      V &     R &   20.05.13 &           840/840 &               4/4 &        DOAO &     This study \\
  &         &             &              &        &      V &     kp1496 &   21.06.25 &           360/900 &               4/4 &        MDM &                \\
8 &  116134 &  09:14:24.8 &  +11:56:25.6 &  0.031 &   kp1494 &  kp1497 &   20.01.08 &           600/600 &               4/4 &         MDM &     This study \\
9$^a$ &   45989 &  16:29:38.4 &  +38:41:39.3 &  0.036 &      V &     R &   20.06.01 &           840/840 &               5/5 &        DOAO &    This study \\
  &         &             &              &        &      V &    kp1497 &   21.06.29 &           360/720 &               3/4 &        MDM &               \\
11 &   59182 &  10:29:11.5 &  +39:06:53.6 &  0.026 &   kp1494 &  kp1496 &   20.01.08 &           600/600 &               4/4 &         MDM &     This study \\
14$^a$ &  125613 &  16:24:51.3 &  +19:25:35.7 &  0.036 &      V &     R &   20.06.01 &           840/840 &               5/5 &        DOAO &     This study \\
   &         &             &              &        &      V &      kp1497 &   21.08.02 &           360/600 &               4/4 &        MDM &                \\
20 &   15709 &  08:04:31.1 &  +40:12:21.8 &  0.040 &   kp1494 &  kp1498 &   20.01.04 &           600/600 &               6/6 &         MDM &     This study \\
23 &    9576 &  08:01:42.6 &  +42:00:19.5 &  0.032 &   kp1494 &  kp1497 &   20.01.04 &           600/600 &               6/6 &         MDM &     This study \\
27 &  125318 &  09:54:18.2 &  +47:17:25.2 &  0.033 &      V &  kp1497 &   20.02.15 &          360/1200 &               5/5 &         MDM &    RGG \\
29$^a$ &   18913 &  15:34:25.6 &  +04:08:06.7 &  0.040 &      V &     R &   20.06.07 &           840/840 &               3/3 &        DOAO &    RGG \\
   &         &             &              &        &      V &     kp1498 &   21.06.28 &           360/900 &               4/4 &        MDM &     \\
30 &  109016 &  10:14:40.2 &  +19:24:49.0 &  0.029 &      V &  kp1496 &   20.03.04 &          360/1200 &               4/4 &         MDM &    RGG \\
31 &   12793 &  10:51:00.7 &  +65:59:40.5 &  0.033 &      V &  kp1497 &   20.03.04 &          360/1200 &               4/4 &         MDM &    RGG \\
32 &   91579 &  12:03:25.7 &  +33:08:46.2 &  0.035 &      V &  kp1497 &   20.03.04 &          360/1200 &               5/5 &         MDM &    RGG \\
34$^a$ &   99052 &  16:05:31.9 &  +17:48:26.2 &  0.032 &      V &     R &   20.06.07 &           840/840 &               3/3 &        DOAO &    RGG \\
   &         &             &              &        &      V &     kp1497 &   21.08.02 &           360/600 &        4/4 &        MDM &        \\
35 &  112250 &  11:23:15.8 &  +24:02:05.2 &  0.025 &      V &  kp1496 &   20.03.04 &          360/1200 &               4/4 &         MDM &    RGG \\
36 &   47066 &  08:51:25.8 &  +39:35:41.8 &  0.041 &      V &  kp1498 &   20.02.15 &          360/1200 &               4/5 &         MDM &    RGG \\
38$^a$ &  104565 &  13:43:32.1 &  +25:31:57.7 &  0.029 &      V &     R &   20.05.13 &           840/840 &               4/4 &        DOAO &    RGG \\
   &         &             &              &        &      V &  kp1496 &   21.06.25 &           360/900 &               3/3 &        MDM &        \\
40$^a$ &   79874 &  15:26:37.4 &  +06:59:41.7 &  0.038 &               V &             R &   20.06.07 &           840/840 &               3/3 &        DOAO &    RGG \\
   &         &             &              &        &      V &     kp1497  &   21.06.28 &           360/900 &               5/5 &        MDM &     \\
\enddata
\tablecomments{(1) Identification number assigned in this study. (2) NASA-Sloan Atlas identification number. (3) Right Ascension in J2000.0. (4) Declination in J2000.0. (5) Redshift. (6) Filter for the continuum part observation. (7) Filter for the $\mathrm{H}\alpha$ part observation. (8) Observation date. (9) Exposure time in units of s for the continuum part and $\mathrm{H}\alpha$ part observation. (10) The number of the epochs for the continuum part and $\mathrm{H}\alpha$ part observation. (11) Observation site. (12) Name of the original sample. This study: IMBH candidates identified in this study. RGG: IMBH candidates selected from \citealt{Reines+13}.\\
$^a$Targets observed at both observatories.
}
\end{deluxetable*}
\end{longrotatetable}
%%%%%%%%%%%%%%%%%%%%%%%%%%%%%%%%%%%%%%%%%%%%%%%%%%%%%%%%%%%%%%%%%%%%%%%%%%%%%%%

\section{observation and data reduction\label{sec:observation}}

\subsection{Observations}
%%% FIGURE %%%%%%%%%%%%%%%%%%%%%%%%%%%%%%%%%%%%%%%%%%%%%%%%%%%%%%%%%%%%%%%%%%%%
\begin{figure}
\centering
\includegraphics[width=80mm]{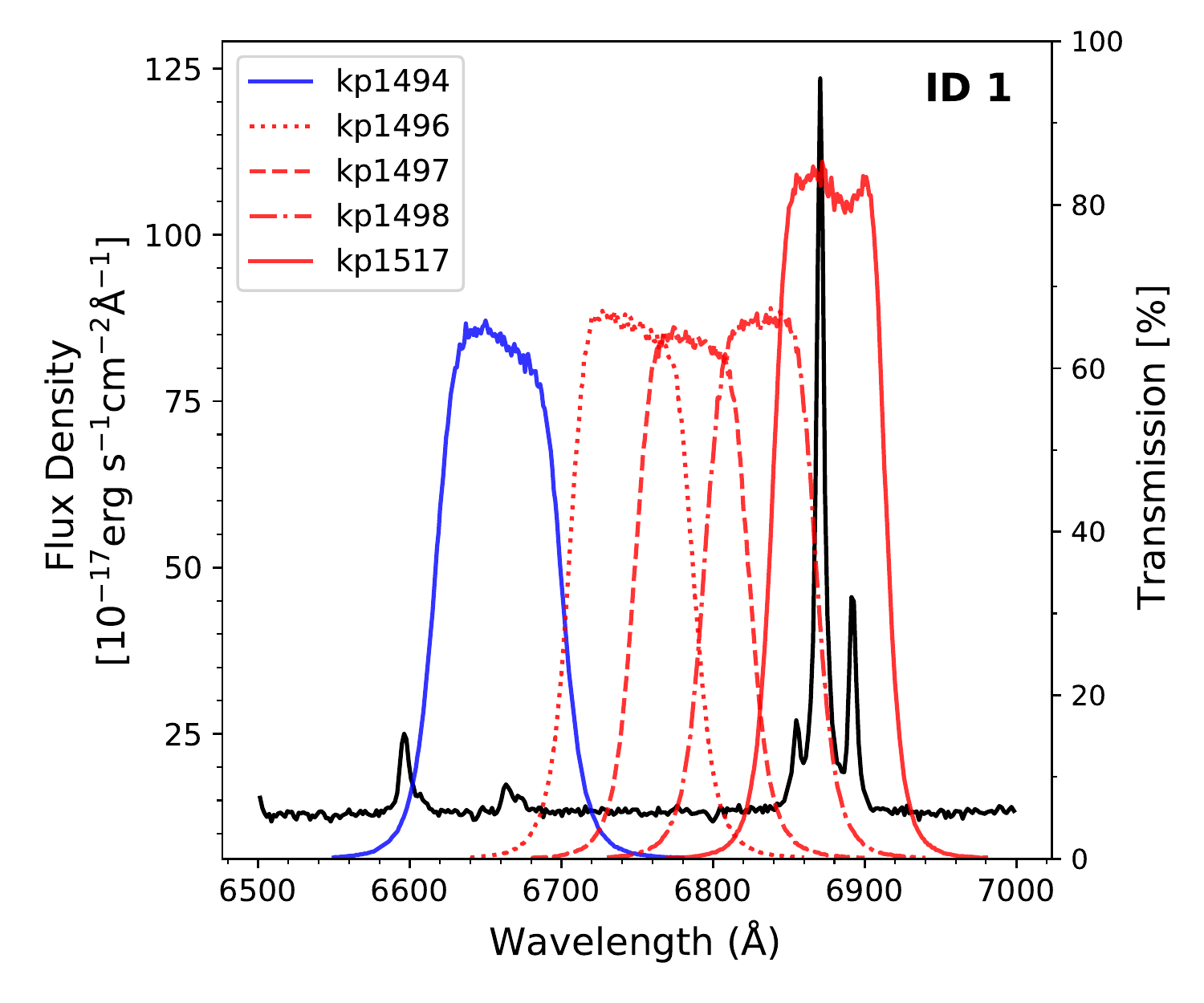}
\caption{Response function of the selected filters. As an example, the SDSS spectrum of ID1 is plotted (black) along with the kp1494 filter for obtaining continuum flux (blue)
and the kp1517 filter for \Ha. The response function of three additional filters (kp1496, kp1497, and kp1498) are also presented. Note that depending on the redshift of each target, we used a different set of filters. 
   \label{fig:res_ex}}
\end{figure}
%%%%%%%%%%%%%%%%%%%%%%%%%%%%%%%%%%%%%%%%%%%%%%%%%%%%%%%%%%%%%%%%%%%%%%%%%%%%%%%

\begin{figure*}
\centering
\includegraphics[angle=0,width=160mm]{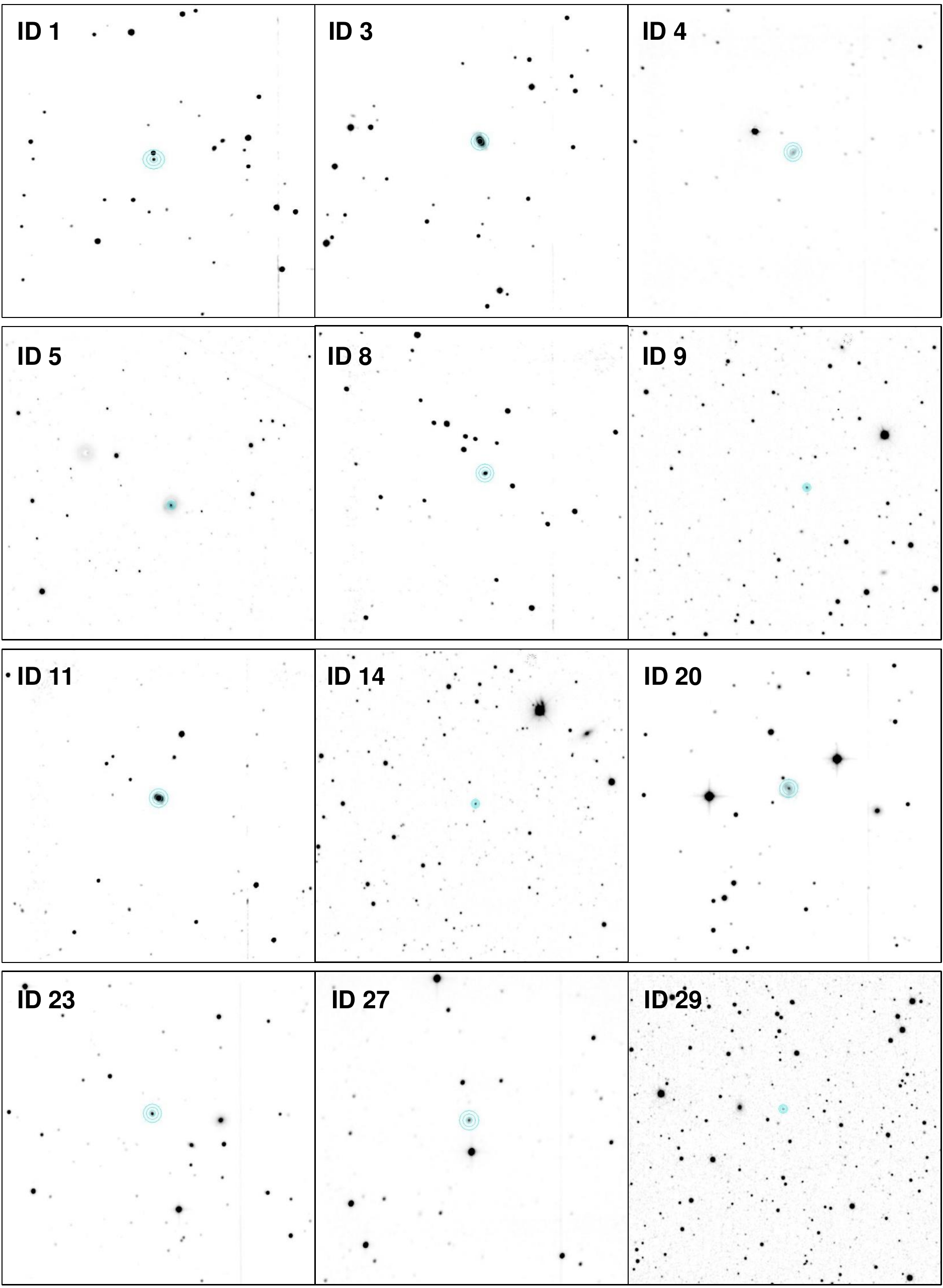}
\caption{Images of the 20 candidates observed with the H$\alpha$ focused filter (see Table \ref{tab:table_obs}). The upper direction is the north, and the left is the east. The 5", 10", 15" apertures are denoted with cyan circles. The ID assigned in this study is shown in the upper left corner of each panel.   \label{fig:fov1}}
\end{figure*}

\begin{figure*}
\centering
\addtocounter{figure}{-1}
\includegraphics[angle=0,width=160mm]{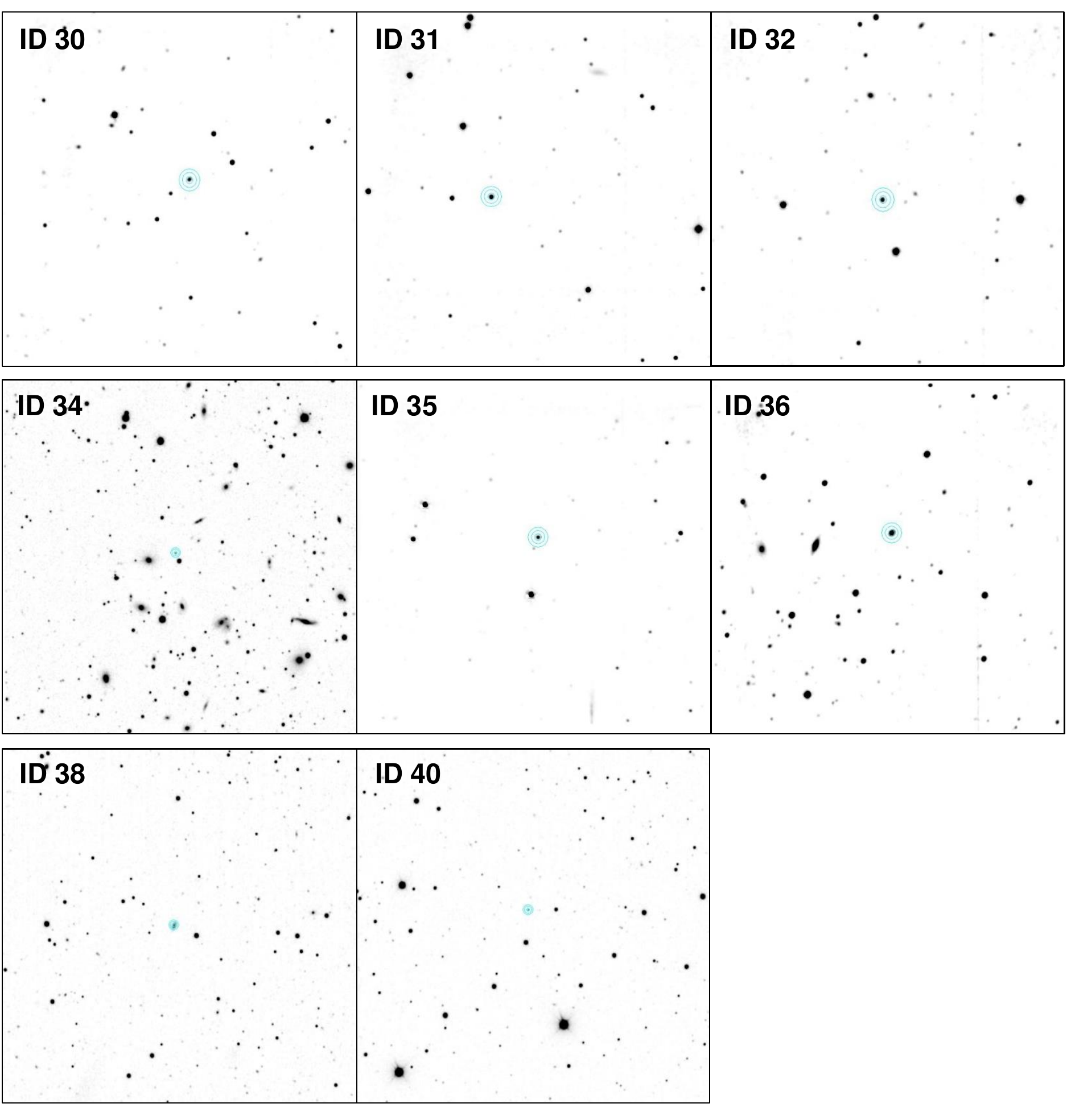}
\caption{Continued \label{fig:fov2}}
\end{figure*}

For an intra-night variability test we observed 20 targets out of 40 IMBH candidates in 2020 and 2021. These targets have expected time lag $<$ 2.4 days, and they were observable during our monitoring runs. In 2020, from 4th January to 7th June, we observed 13 targets using the 1.3m telescope at the Michigan-Dartmouth-MIT (MDM) observatory on Kitt peak, Tucson, Arizona, USA, with the Templeton CCD, which has 1K$\times$1K pixels with a 0.51$^{\prime\prime}$ pixel size and a 8.49$^{\prime}\times$8.49$^{\prime}$ field of view (FOV). The remaining 7 targets were observed with the 1m telescope at the Deokheung optical astronomical observatory (DOAO) in Jeollanam-do, Korea. We used 4K$\times$4K FLI CCD with 2$\times$2 pixel binning, which provided 0.46$^{\prime\prime}$ pixel size and a 15.8$^{\prime}\times$15.8$^{\prime}$ FOV. In addition, we repeated photometry observations for these 7 objects at the MDM observatory in order to obtain narrow-band filter observations, which were not possible at DOAO due to the lack of narrow filters. More details of the observing facilities can be found in our on-going monitoring studies \citep[e.g.,][]{Woo+19b}. 

To investigate the variability, we targeted 2-4 candidates in each night and carried out repeated observations of 3-6 epochs with a $\sim$1-2 hours cadence. Among available filters,  we used two filters: one filer is selected for obtaining continuum flux without including the \Ha\ emission line, and the other filter is set for the H$\alpha$ emission line (see Figure \ref{fig:res_ex}). Table~\ref{tab:table_obs} lists the observational details of individual targets, which are sorted by their \mbh. 

\subsection{Data reduction}
%%% FIGURE %%%%%%%%%%%%%%%%%%%%%%%%%%%%%%%%%%%%%%%%%%%%%%%%%%%%%%%%%%%%%%%%%%%%
\begin{figure*}
\centering
\includegraphics[angle=0,width=175mm]{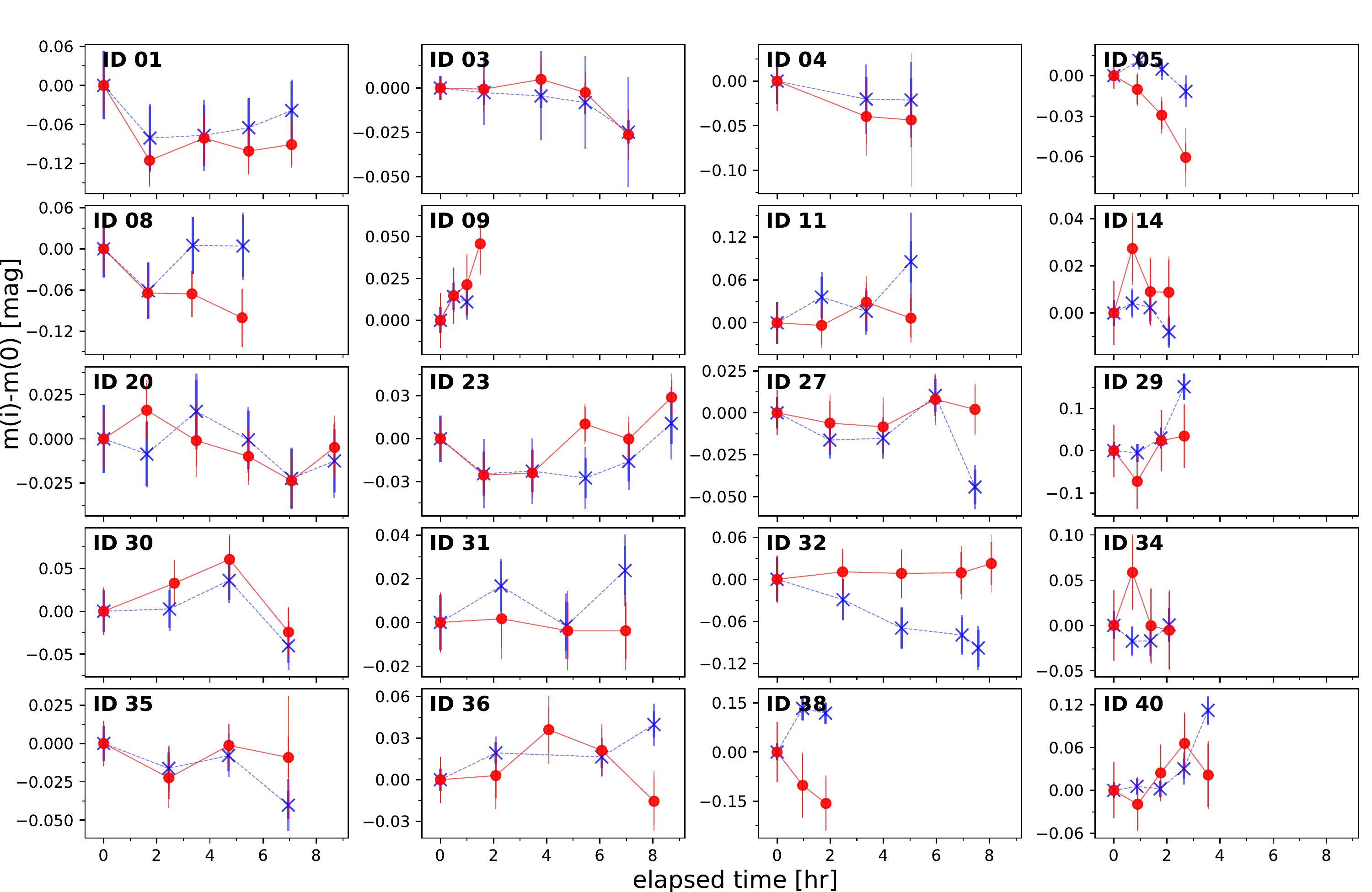}
\caption{Continuum (blue) and H$\alpha$ (red) light curves of the 20 candidates. The total error (thin line) is composed of measurement error (thick line) and the systematic error from differential photometry. 
Note that for H$\alpha$ light curves, we only present the photometry results with narrow-band filters.
The ID assigned in this work is shown in the upper left corner of the panel. \label{fig:light}}
\end{figure*}
%%%%%%%%%%%%%%%%%%%%%%%%%%%%%%%%%%%%%%%%%%%%%%%%%%%%%%%%%%%%%%%%%%%%%%%%%%%%%%%

The standard image preprocess was carried out using the data reduction routine of the Seoul National University AGN monitoring project (SAMP) (\citealt{Woo+19b}; \citealt{Rakshit+19}) as outlined below. First, we performed dark, bias image subtraction and flat-fielding using IRAF routines \citep{Tody86}. 
We then removed cosmic rays using LA-cosmic \citep{vanDokkum01}, 
and derived the astrometric solution with astrometry-net \citep{Lang+10}. Utilizing the solution, we combined the preprocessed exposures of each target using SWarp \citep{Bertin+02} with a median combine method for a deeper single image. Finally, to match the seeing conditions of individual epochs, we performed seeing convolution with \texttt{psf} tool of python \texttt{photutil} package and \texttt{convolution} tool of python \texttt{astropy} package (\citealt{Astropy13}; \citealt{Astropy18}). The preprocessed images of 20 observed candidates are presented in Figure \ref{fig:fov1}.

\subsection{Differential photometry}

We performed differential photometry using the preprocessed images as we aim at detecting flux variation for this study. 
To determine the optimal aperture size for the photometry, we utilized a star in the image brighter than the target, and determined the aperture size for obtaining the highest S/N. We employed the same aperture for all epochs as seeing of each image was already matched.
The typical seeing was FHWM= $\sim$3.3$^{\prime\prime}$ at DOAO and $\sim$2.8$^{\prime\prime}$ at MDM in 2020, and $\sim$2.2$^{\prime\prime}$ at MDM in 2021. We adopted an aperture size $\sim$9$^{\prime\prime}$ for both MDM and DOAO data, which roughly corresponds to $\sim$3 times of the seeing size.

We measured the instrumental magnitude of each object in the FOV with SExtractor \citep{Bertin&Arnouts96}, and constructed the differential light curves utilizing 5 comparison stars. To select the non-variable comparison stars, we first selected 20 brightest stars in the image and performed a variability test. We calculated the magnitude difference ($\Delta$mag) by subtracting the magnitude of the first epoch from the magnitude of each epoch for individual stars. Then, we determined the error weighted average and standard deviation of $\Delta$mag. Based on this calculation, we excluded the stars whose $\Delta$mag is larger than the average by more than a factor of two of the standard deviation. After that we tested each case by randomly selecting 5 stars from the remaining comparison stars and picked a set of 5 stars, whose standard deviation of their $\Delta$mag is the smallest. 
To estimate the uncertainty of the calibrated magnitude in the light curves, we combined the measurement error ($\sigma_{m}$)
and the standard deviation of the $\Delta$mag of the selected 5 comparison stars ($\sigma_{s}$) in quadrature: 
\begin{equation}
\label{eq:err}
\sigma = \sqrt{{\sigma_{m}}^2+{\sigma_{s}}^2}.
\end{equation}
The calibrated light curves from differential photometry are presented in Figure \ref{fig:light}.

\section{results and analysis\label{sec:variability}}

\subsection{Variability amplitude\label{sec:vari_amp}}

To investigate the variability of the observed IMBH candidates, we calculated three different parameters, namely, the flux ratio between the maximum and minimum brightness ($R_{max}$), standard deviation of the magnitudes in the light curve (RMS), and the normalized excess variance of the flux ($F_{var}$) defined as
\begin{equation}
\label{eq:fvar}
F_{var} = \frac{\sqrt{f_{\sigma}^2-f_{\delta}^2}}{f_{avg}}
\end{equation}
where $f_{avg}$ is the average of the flux of the target, $f_{\sigma}$ is the standard deviation of the flux, and $f_{\delta}$ is the average of the flux uncertainty at each epoch (\citealt{Rodriguez-Pascual+97}, \citealt{Walsh+09}). These measurements of the variability amplitude are presented in Table \ref{tab:table_vari}.
Note that $F_{var}$ is often considered as the most reliable parameter of variability because it is accounted for the uncertainty of the flux measurement.

We examined the variability of AGN continuum and \Ha\ emission line using the two light curves presented in Section \ref{sec:observation}.
Note that since the flux in the \Ha\ light curve, which was measured with the adopted narrow or broad band filter, contains \Ha\ emission line flux as well as 
AGN continuum and non-variable host galaxy contribution. In the case of the continuum light curve, the measured flux is the sum of the flux mainly from AGN continuum 
and host galaxy (see Figure \ref{fig:res_ex}). Thus, the measured variability amplitude depends on the dominance of the AGN component over the host galaxy flux. In other words, even if AGN is highly variable, non-variable host galaxy contribution may dilute the flux variability in the light curve. 

The measured R$_{max}$ in the continuum light curve ranges from 1.01 to 1.33, indicating that the maximum intra-night variability is over 0.3 magnitude, while the highest and the median value of RMS are 0.13 mag and 0.01 mag, respectively. The largest RMS is similar to the median RMS $\sim$0.14 mag of typical Seyfert 1 galaxies \citep{Rakshit&Stalin17}. However, several targets in the sample show very weak variability, leading to relatively low median RMS. If we assume that the measurement error is $\sim$1$\%$, continuum flux variability is detected for all targets, with a typical RMS variability $\gtrsim$ 0.01 mag. Considering the measurement uncertainty of individual targets, we identified 9 candidates (namely, IDs 5, 27, 29, 30, 32, 35, 36, 38, and 40) as more secure variable targets with a measurable $F_{var}$ of $\sim$0.01-0.05 mag, which suggests that the detected variability is a few \%. The other 11 targets have no measurable $F_{var}$ value, which may be caused by relatively large measurement error compared to the intrinsic variability, dilution due to the dominance of the non-variable host galaxy flux, or the lack of intrinsic AGN variability. 

In comparison, the measured R$_{max}$ in the \Ha\ light curve ranges from 1.01 to 1.16,
while RMS shows a narrower range of $\sim$0.01-0.06 mag than that of the continuum light curve. The median values of $R_{max}$, RMS, and $F_{var}$ are 1.04, 0.02 mag, and 0.01 mag, respectively, which is similar to the median values from the continuum light curve. Whereas their highest values are 1.16, 0.06 mag, and 0.02 mag, respectively, showing that the variability of \Ha\ is less significant than continuum. As all targets except for ID 31 show RMS $\gtrsim$ 0.01 mag, the variability of \Ha\ seems detected for most tagets. However, after considering the measurement error, we only obtained excess variance for 4 targets (namely, IDs 1, 5, 23 and 30) with $F_{var}$ $\sim$0.01-0.02 mag.

To better understand the measured variability amplitude in the \Ha\ light curve, we performed spectral analysis to measure the fraction of the broad \Ha\ component in the total flux obtained through each filter. By multiplying the response function of the used filter to the SDSS spectrum of each object, we calculated the fraction of the broad \Ha. 
Note that the calculated fraction is an upper limit since the adopted aperture of $\sim$9$^{\prime\prime}$ in photometry is much larger than the $\sim$3$^{\prime\prime}$ fiber used for the SDSS spectra. 
For the broad R filter, the fraction of the broad \Ha\ is insignificant ($\lesssim$1$\%$), indicating that the detection of the \Ha\ variability is very challenging even if the broad \Ha\ emission line intrinsically varies by $<$100\%. In contrast, the broad \Ha\ fraction in the narrow filter is significantly larger (2 -- 24\%; see last column in Table \ref{tab:table_vari}), which is much larger than the measurement uncertainty of 1\%. Thus, if the intrinsic variability is significant, it is likely that the variability of \Ha\ is detected in the narrow-filter based light curves.

For example, IDs 1, 5, 23, and 30 from MDM observation have considerable \Ha\ fraction (14.1$\%$, 1.9$\%$, 4.5$\%$, and 3.5$\%$), and the excess variance is also detected as $F_{var}$ $\sim$1-2$\%$.  Other targets with no detection of excess variance may have a very small flux ratio of the \Ha\ emission or weak variability. Note that for ID 1 and ID 23, we obtained no excess variance $F_{var}$ in the continuum light curve, which is likely due to the fact that non-variable host galaxy component is dominant in the continuum light curve, or that the continuum flux variability amplitude is smaller than the measurement error. 

In contrast, the targets observed with broad R filter (namely, IDs 9, 14, 29, 34, 38, and 40) have no measurable $F_{var}$ value due to the insignificant \Ha\ fraction. The dominant continuum may have diluted the intrinsic \Ha\ variability if any. 
In the case of ID5 observed with a broad R filter, we obtained somewhat unexpected result. Since the fraction of \Ha\ is less than 0.1$\%$ in the total flux measured with photometry. the 1\% excess variance seems too high to be detected because the \Ha\ flux needed to vary by a factor of $>$10. Note that its AGN continuum variability is negligible as we detect no excess variance in the continuum light curve. We consider a possibility that the measurement error in the \Ha\ light curve is underestimated and smaller than that of continuum light curve. 

In summary, we detected RMS variability of all 20 targets, and obtained excess variance $F_{var}$ for 9 targets based on the observed intra-night continuum light curves. In the case of the \Ha\ light curves, we determined RMS variability of $\gtrsim$1\% for all targets except for ID 31, and reliable excess variance for 4 targets. Only two candidate, IDs 5 and 30, showed reliable excess variance $F_{var}$ in the light curve of both continuum and H$\alpha$, suggesting that they are the best IMBH candidate in the sample. The other 9 targets, namely, IDs 1, 23, 27, 29, 32, 35, 36, 38, and 40, are also good candidates for further studies since they showed excess variance either in the continuum or \Ha\ light curve.

%%% FIGURE %%%%%%%%%%%%%%%%%%%%%%%%%%%%%%%%%%%%%%%%%%%%%%%%%%%%%%%%%%%%%%%%%%%%
\begin{figure*}[!htbp]
\centering
\includegraphics[angle=0,width=0.95\textwidth]{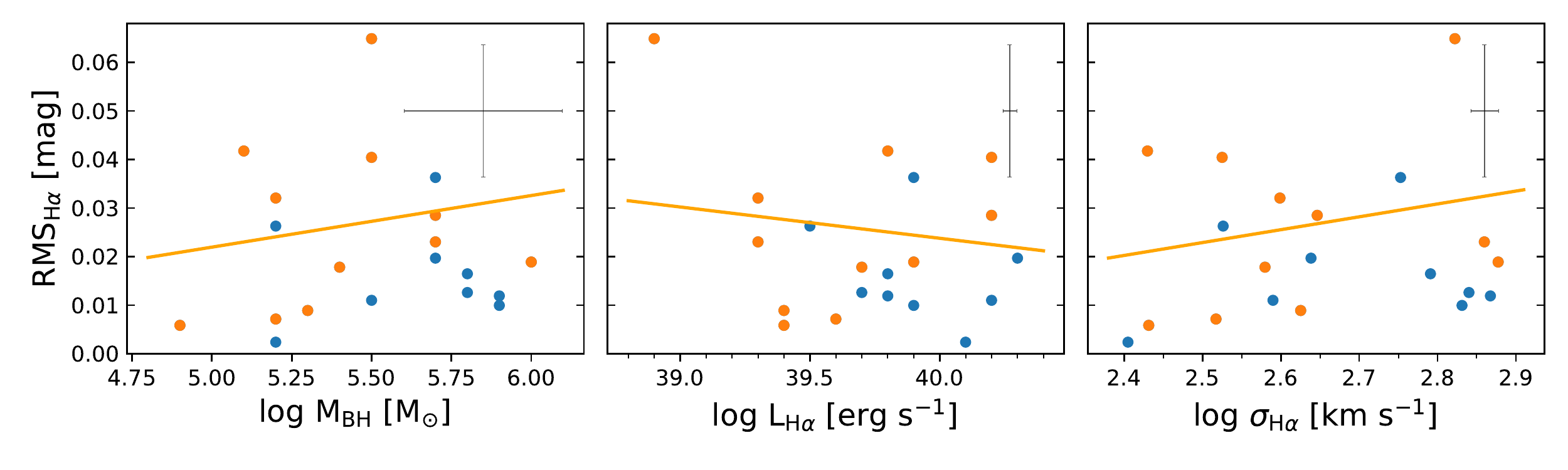}
\caption{The correlations of RMS$_{\rm H\alpha}$ with M$_{\rm BH}$, L$_{\rm H\alpha}$, and $\sigma_{\rm H\alpha}$. The 11 candidates having large variability amplitude selected in Section \ref{sec:vari_amp} are plotted in orange, and the other candidates with weak variability are plotted in blue. A typical error is shown in the upper right corner. The solid lines show the best linear fit for the 11 variable candidates. } \label{fig:agn_vari}
\end{figure*}
%%%%%%%%%%%%%%%%%%%%%%%%%%%%%%%%%%%%%%%%%%%%%%%%%%%%%%%%%%%%%%%%%%%%%%%%%%%%%%%

\begin{longrotatetable}
\begin{deluxetable*}{llllllllllllll}%[!hptb]
\tablecaption{Variability information of the observed sample \label{tab:table_vari}}
\tablecolumns{14}
\tablehead
{
\colhead{ID} & \colhead{$\log M_\mathrm{BH}$} & \colhead{$\log L_{\mathrm{H}\alpha}$} & \colhead{$(1+z)\,\tau$} & \colhead{$\sigma_{\mathrm{H}\alpha}$}  & \colhead{$\mathrm{RMS}_\mathrm{cont.}$} & \colhead{$R_\mathrm{max;\,cont.}$} & \colhead{$F_\mathrm{var;\,cont.}$} & \colhead{$\mathrm{RMS}_{\mathrm{H}\alpha}$} & \colhead{$R_{\mathrm{max;\,H}\alpha}$} & \colhead{$F_{\mathrm{var;\,H}\alpha}$} & \colhead{Filter\_cont}& \colhead{Filter\_$\mathrm{H}\alpha$}& \colhead{$f_{\mathrm{H}\alpha}$/$f_{tot.}$} \\
%& & & &  &  &  &  & & & \\
\colhead{(1)} & \colhead{(2)} & \colhead{(3)} & \colhead{(4)} & \colhead{(5)} & \colhead{(6)} & \colhead{(7)} & \colhead{(8)} & \colhead{(9)} & \colhead{(10)} & \colhead{(11)} & \colhead{(12)}& \colhead{(13)}& \colhead{(14)}  \\
}
\startdata
1 &    5.5 &   40.2 &    2.4 &  335 &     0.03 &      1.08 &         - &  0.04 &  1.11 &  0.01 &   kp1494  & kp1517 &        14.1 \\
3 &    5.5 &   40.2 &    2.4 &  389 &     0.01 &      1.02 &         - &  0.01 &  1.03 &     - &   kp1494  & kp1497 &         5.1 \\
4 &    5.7 &   40.3 &    2.7 &  435 &     0.01 &      1.02 &         - &  0.02 &  1.04 &     - &   kp1494  & kp1497 &        10.3 \\
5$^a$ &    5.7 &   39.3 &    0.9 &  724 &    $<$ 0.01 &      1.01 &      - &  0.01 &  1.02 &  0.01 &      V    &      R &         0.1 \\
  &        &        &        &      &     0.01 &      1.02 &  $<$ 0.01 &  0.02 &  1.06 &  0.02 &      V    & kp1496 &         1.9 \\
8 &    5.7 &   39.9 &    1.7 &  566 &     0.03 &      1.06 &         - &  0.04 &  1.10 &     - &   kp1494  & kp1497 &         9.3 \\
9$^a$ &    5.8 &   39.8 &    1.6 &  618 &     0.01 &      1.02 &         - &  0.01 &  1.02 &     - &      V    &      R &         0.4 \\
  &        &        &        &      &     0.01 &      1.01 &         - &  0.02 &  1.04 &     - &      V    & kp1497 &         7.3 \\
11 &    5.8 &   39.7 &    1.5 &  692 &     0.03 &      1.08 &         - &  0.01 &  1.03 &     - &   kp1494 &  kp1496 &         6.2 \\
14$^a$ &    5.9 &   39.9 &    1.7 &  678 &     0.01 &      1.04 &         - &  0.01 &  1.03 &     - &      V   &       R &         0.3 \\
 &        &        &        &   &     $<$ 0.01 &      1.01 &    - &  0.01 &  1.03 &     - &       V   & kp1497 &         3.5 \\
20 &    5.9 &   39.8 &    1.6 &  737 &     0.01 &      1.04 &         - &  0.01 &  1.04 &     - &   kp1494 &  kp1498 &         4.5 \\
23 &    6.0 &   39.9 &    1.7 &  754 &     0.01 &      1.04 &         - &  0.02 &  1.05 &  0.01 &   kp1494 &  kp1497 &         4.5 \\
27 &    4.9 &   39.4 &    1.0 &  270 &     0.02 &      1.05 &      0.01 &  0.01 &  1.02 &     - &      V   &  kp1497 &         3.0 \\
29$^a$ &    5.1 &   39.8 &    1.6 &  269 &     0.13 &      1.33 &      0.02 &  0.01 &  1.03 &     - &      V   &       R &         1.4 \\
   &        &        &        &      &     0.06 &      1.15 &      0.05 &  0.04 &  1.10 &     - &      V   &  kp1498 &        23.6 \\
30 &    5.2 &   39.3 &    0.9 &  397 &     0.03 &      1.07 &      0.01 &  0.03 &  1.08 &  0.01 &      V   &  kp1496 &         3.5 \\
31 &    5.2 &   40.1 &    2.1 &  254 &     0.01 &      1.02 &         - &  $<$ 0.01 &  1.01 & - &      V   &  kp1497 &         6.8 \\
32 &    5.2 &   39.6 &    1.2 &  329 &     0.04 &      1.09 &      0.02 &  0.01 &  1.02 &     - &      V   &  kp1497 &         7.6 \\
34$^a$ &    5.2 &   39.5 &    1.1 &  336 &     0.04 &      1.09 &         - &  0.04 &  1.09 &     - &      V   &       R &         0.6 \\
   &        &        &        &      &     0.01 &      1.02 &         - &  0.03 &  1.06 &     - &      V   &  kp1497 &        11.5 \\
35 &    5.3 &   39.4 &    1.0 &  422 &     0.02 &      1.04 &      $<$ 0.01 &  0.01 &  1.02 & - &      V   &  kp1496 &         3.0 \\
36 &    5.4 &   39.7 &    1.4 &  380 &     0.01 &      1.04 &      0.01 &  0.02 &  1.05 &     - &      V   &  kp1498 &         6.8 \\
38$^a$ &    5.5 &   38.9 &    0.6 &  664 &     0.04 &      1.11 &      0.02 &  0.02 &  1.04 &     - &      V   &       R &         0.3 \\
   &        &        &        &      &     0.06 &      1.13 &      0.05 &  0.06 &  1.16 &     - &      V   &  kp1496 &         6.2 \\
40$^a$ &    5.7 &   40.2 &    2.3 &  443 &     0.01 &      1.03 &         - &  0.02 &  1.04 &     - &      V   &       R &         1.1 \\
   &        &        &        &      &     0.04 &      1.11 &      0.04 &  0.03 &  1.08 &     - &      V   &  kp1497 &        15.3 \\
\enddata
\tablecomments{(1) Identification number assigned in this study. (2) Estimated black hole mass in units of $ \rm M_{\odot}$. (3) Luminosity of the H$\alpha$ broad component in units of erg s$^{-1}$. (4) Expected time lag in units of days. (5) Velocity dispersion of the H$\alpha$ broad component in units of km s$^{-1}$. (6) RMS of the continuum part light curve in units of mag. (7) $R_{max}$ of the continuum part light curve. (8) $F_{var}$ of the continuum part light curve in units of mag. (9) RMS of the H$\alpha$ part light curve in units of mag. (10) $R_{max}$ of the H$\alpha$ part light curve. (11) $F_{var}$ of the H$\alpha$ part light curve in units of mag. (12) Filter for the continuum part observation. (13) Filter for the H$\alpha$ part observation. (14) Flux ratio of the broad \Ha\ line to the total flux observed with each filter in units of $\%$.\\
$^a$Observed at both observation sites.
}
\end{deluxetable*}
\end{longrotatetable}

%%%%%%%%%%%%%%%%%%%%%%%%%%%%%%%%%%%%%%%%%%%%%%%%%%%%%%%%%%%%%%%%%%

\subsection{Variability vs. AGN properties}

We compared the AGN properties and RMS$_{\rm H\alpha}$ measured with narrow-filter based light curves (see Figure \ref{fig:agn_vari}). We computed a least-square linear regression for the highly variable candidates identifed in the previous section using \texttt{linregress} tool of python \texttt{scipy} package \citep{Virtanen+20}. The variability amplitude shows a weak positive correlation with M$_{\rm BH}$ and $\sigma_{H\alpha}$ , whose correlation coefficients (r-value) are 0.19, 0.24 and two sided p-values for a null hypothesis of no correlation are 0.58, 0.48, respectively. On the other hand, L$_{H\alpha}$ is anti-correlated with the RMS$_{\rm H\alpha}$ having weak correlation coefficients (r-value = -0.14 and p-value = 0.67). Similar results have been found by \citet{Rakshit&Stalin17}, who reported that the variability strength has an anti-correlation with AGN luminosity, but a correlation with the M$_{\rm BH}$ and velocity dispersion of the broad permitted lines. The correlation of their sample at z$<$0.2 is not significant (i.e., p-value $>$ 0.1), suggesting that our result is not inconsistent with that of \citet{Rakshit&Stalin17}.
However, the time-based line and the number of epoch are clearly limited, and further studies with sufficient data are required to unveil the nature of the variability
correlation of IMBH candidates. 

\section{discussion\label{sec:dis}}

Several previous studies searched for low mass BHs hosted in the dwarf galaxies by detecting the spectral signature of BH accretion using the SDSS galaxy sample, and identified IMBH candidates with broad H$\alpha$ emission line (e.g., \citealt{Reines+13}; \citealt{Baldassare+15}; \citealt{Eun+17}; \citealt{Chilingarian+18}). We compared the selection criteria and the results of the previous works with this study. For example, \citet{Eun+17} found 611 broad H$\alpha$ candidates from SDSS DR7 data. Their mean value of H$\alpha$ luminosity ($10^{40.7\pm 0.3}$ erg s$^{-1}$) is larger than the mean value of our sample of 131 objects with a broad H$\alpha$ ($10^{40.3}$ erg s$^{-1}$). Velocity dispersion of H$\alpha$ of most of their targets is also larger than 1000 km s$^{-1}$ (see Figure 10 in \citealt{Eun+17}), while $\sim70\%$ of our broad H$\alpha$ candidates (92/131) has $\sigma_{H\alpha}$ smaller than 1000 km s$^{-1}$. Consequently, the estimated M$_{\rm BH}$ of the sample in \citet{Eun+17} is larger than $10^6  \rm M_{\odot}$. In contrast, we have found 25 targets whose estimated M$_{\rm BH}$ is less than $10^6  \rm M_{\odot}$. Note that \cite{Eun+17} conservatively selected clear type 1 AGN candidates in order to avoid the false detection. In contrast, we set the lower $\sigma_{H\alpha}$ cut for the definition of the type 1 AGN, and performed more delicate spectral analysis using $\chi_{red}^2$ and BIC to find the targets with the weak H$\alpha$ broad component.

Similar comparison can be done with the broad H$\alpha$ candidates from \citet{Reines+13}. They have identified the candidates with dimmer H$\alpha$ broad component than ours as their mean $L_{H\alpha}$ ($10^{39.7}$ erg s$^{-1}$) is lower than that of our 131 broad H$\alpha$ candidates (i.e., $10^{40.3}$ erg s$^{-1}$). Also, their mean $\sigma_{H\alpha}$ ($\sim$670 km s$^{-1}$) is lower than ours ($\sim$870 km s$^{-1}$). The main difference comes from the fact that they only surveyed dwarf galaxies at z $<$ 0.055, by imposing stellar mass limit, i.e, $<3\times10^9$ $ \rm M_{\odot}$. Since we searched more distant targets, it is likely that we missed the candidates having a weak H$\alpha$ broad component at z $>$ 0.055. Note that among the 25 IMBH candidates in this study, whose M$_{\rm BH}$ is less than $10^6  \rm M_{\odot}$, only one target is overlapped with the IMBH candidates in \citet{Reines+13}. While both samples were selected from the SDSS catalog based on similar selection criteria,
the details of the selection scheme and the spectral decomposition analysis caused the difference. It shows that more missing IMBH candidates can be identified among SDSS galaxies by refining the selection procedures.

\citet{Chilingarian+18} also performed a similar study to identify IMBHs among SDSS galaxies. They reported IMBH candidates whose estimated M$_{\rm BH}$ is less than $10^5 \rm M_{\odot}$ based on their own non-parametric emission line fitting method along with X-ray flux analysis \citep{Chilingarian+17}. While it is possible that for some of the candidates, their fitting method constructing multiple template sets of flux-normalized Gaussians may suffer an overfitting problem, underestimating the luminosity and velocity dispersion of the broad H$\alpha$ component, they demonstrated that spectral analysis can be utilized to find IMBH candidates even with \mbh\ $\sim10^{4} \rm M_{\odot}$.

The presence of a broad \Ha\ may not be sufficient to be qualified as type 1 AGN since the line width of \Ha\ of IMBH candidates is relatively narrow compared to the conventional limit of $\sim$1,000 \kms , 
and other sources such as supernovae are able to exhibit a broad \Ha\ \citep[e.g.,][]{Izotov+07, Izotov&Thuan09b,Graur&Maoz13}.
However, by detecting variability, these targets can be confirmed as mass accreting IMBHs. While various studies performed variability analysis over a relatively long time scale for identifying IMBH candidates \citep[e.g.,][]{Baldassare+18, Martinez+20}, Short-time scale, i.e., intra-night variability studies are rare. We compare our results with the intra-night variability analysis conducted by \citet{Kim+18}, who monitored AGN candidates in the SMBH regime.  While the FOV ($\sim$10$^{\prime}\times$10$^{\prime}$) of our photometric observations is much smaller and the time baseline is shorter (i.e., $\sim$8 hours) than those of \cite{Kim+18}, we obtained measurable variability amplitudes for a subsample of the IMBH candidates, suggesting that it is possible to detect variability of IMBH candidates based on intra-night monitoring programs with a relatively short cadence. Future narrow-band monitoring programs with well defined strategies on the time base line and cadence can provide valuable assessment and confirmation of IMBH candidates.

\section{summary and conclusion \label{sec:con}}

To detect the accretion signature of IMBH candidates, we identified the galaxies with a broad H$\alpha$ emission line, and performed further analysis to constrain the nature of these targets. The main results are summarized as follows.
\begin{itemize}
\item We performed spectral analysis in the H$\alpha$ emission line region using a large sample of the SDSS DR7 galaxy, and newly found 131 targets with a weak broad H$\alpha$ line. Among them, 25 targets have M$_{\rm BH}$ less than $10^6 \rm M_{\odot}$.

\item To find additional evidence of AGN, we obtained X-ray data and measured X-ray luminosity for 19 candidates. However the estimated M$_{\rm BH}$ of these targets is larger than $10^6 \rm M_{\odot}$.

\item We observed 20 IMBH candidates for an intra-night variability test and reported 11 objects with the largest variability amplitude as the best IMBH candidates.
\end{itemize}

Our study demonstrated that an intensive monitoring campaign with a larger-aperture telescope along with a narrow-band filter can provide strong constraints over the population of IMBHs at low z, by overcoming the relatively large measurement error and the limited number of epochs of this study. 
The best IMBH candidates obtained in this study are one of the best suitable targets for further studies, i.e., spectroscopic reverberation mapping, to confirm them as IMBHs and determine reliable black hole masses.

\acknowledgements{
This work was supported by the National Research Foundation of Korea grant funded by the Korean government (NRF-2021R1A2C3008486).}

\bibliography{ms}{}

\begin{thebibliography}{}
\expandafter\ifx\csname natexlab\endcsname\relax\def\natexlab#1{#1}\fi
\providecommand{\url}[1]{\href{#1}{#1}}
\providecommand{\dodoi}[1]{doi:~\href{http://doi.org/#1}{\nolinkurl{#1}}}
\providecommand{\doeprint}[1]{\href{http://ascl.net/#1}{\nolinkurl{http://ascl.net/#1}}}
\providecommand{\doarXiv}[1]{\href{https://arxiv.org/abs/#1}{\nolinkurl{https://arxiv.org/abs/#1}}}

\bibitem[{{Abazajian} {et~al.}(2009){Abazajian}, {Adelman-McCarthy},
  {Ag{\"u}eros}, {Allam}, {Allende Prieto}, {An}, {Anderson}, {Anderson},
  {Annis}, {Bahcall}, {Bailer-Jones}, {Barentine}, {Bassett}, {Becker},
  {Beers}, {Bell}, {Belokurov}, {Berlind}, {Berman}, {Bernardi}, {Bickerton},
  {Bizyaev}, {Blakeslee}, {Blanton}, {Bochanski}, {Boroski}, {Brewington},
  {Brinchmann}, {Brinkmann}, {Brunner}, {Budav{\'a}ri}, {Carey}, {Carliles},
  {Carr}, {Castander}, {Cinabro}, {Connolly}, {Csabai}, {Cunha}, {Czarapata},
  {Davenport}, {de Haas}, {Dilday}, {Doi}, {Eisenstein}, {Evans}, {Evans},
  {Fan}, {Friedman}, {Frieman}, {Fukugita}, {G{\"a}nsicke}, {Gates},
  {Gillespie}, {Gilmore}, {Gonzalez}, {Gonzalez}, {Grebel}, {Gunn},
  {Gy{\"o}ry}, {Hall}, {Harding}, {Harris}, {Harvanek}, {Hawley}, {Hayes},
  {Heckman}, {Hendry}, {Hennessy}, {Hindsley}, {Hoblitt}, {Hogan}, {Hogg},
  {Holtzman}, {Hyde}, {Ichikawa}, {Ichikawa}, {Im}, {Ivezi{\'c}}, {Jester},
  {Jiang}, {Johnson}, {Jorgensen}, {Juri{\'c}}, {Kent}, {Kessler}, {Kleinman},
  {Knapp}, {Konishi}, {Kron}, {Krzesinski}, {Kuropatkin}, {Lampeitl},
  {Lebedeva}, {Lee}, {Lee}, {French Leger}, {L{\'e}pine}, {Li}, {Lima}, {Lin},
  {Long}, {Loomis}, {Loveday}, {Lupton}, {Magnier}, {Malanushenko},
  {Malanushenko}, {Mandelbaum}, {Margon}, {Marriner}, {Mart{\'\i}nez-Delgado},
  {Matsubara}, {McGehee}, {McKay}, {Meiksin}, {Morrison}, {Mullally}, {Munn},
  {Murphy}, {Nash}, {Nebot}, {Neilsen}, {Newberg}, {Newman}, {Nichol},
  {Nicinski}, {Nieto-Santisteban}, {Nitta}, {Okamura}, {Oravetz}, {Ostriker},
  {Owen}, {Padmanabhan}, {Pan}, {Park}, {Pauls}, {Peoples}, {Percival}, {Pier},
  {Pope}, {Pourbaix}, {Price}, {Purger}, {Quinn}, {Raddick}, {Re Fiorentin},
  {Richards}, {Richmond}, {Riess}, {Rix}, {Rockosi}, {Sako}, {Schlegel},
  {Schneider}, {Scholz}, {Schreiber}, {Schwope}, {Seljak}, {Sesar}, {Sheldon},
  {Shimasaku}, {Sibley}, {Simmons}, {Sivarani}, {Allyn Smith}, {Smith},
  {Smol{\v{c}}i{\'c}}, {Snedden}, {Stebbins}, {Steinmetz}, {Stoughton},
  {Strauss}, {SubbaRao}, {Suto}, {Szalay}, {Szapudi}, {Szkody}, {Tanaka},
  {Tegmark}, {Teodoro}, {Thakar}, {Tremonti}, {Tucker}, {Uomoto}, {Vanden
  Berk}, {Vandenberg}, {Vidrih}, {Vogeley}, {Voges}, {Vogt}, {Wadadekar},
  {Watters}, {Weinberg}, {West}, {White}, {Wilhite}, {Wonders}, {Yanny},
  {Yocum}, {York}, {Zehavi}, {Zibetti}, \& {Zucker}}]{Abazajian+09}
{Abazajian}, K.~N., {Adelman-McCarthy}, J.~K., {Ag{\"u}eros}, M.~A., {et~al.}
  2009, \apjs, 182, 543, \dodoi{10.1088/0067-0049/182/2/543}

\bibitem[{Abbott {et~al.}(2020)Abbott, Abbott, Abraham, Acernese, Ackley,
  Adams, Adhikari, Adya, Affeldt, Agathos, Agatsuma, Aggarwal, Aguiar, Aich,
  Aiello, Ain, Ajith, Akcay, Allen, Allocca, Altin, Amato, Anand, Ananyeva,
  Anderson, Anderson, Angelova, Ansoldi, Antier, Appert, Arai, Araya, Areeda,
  Ar\`ene, Arnaud, Aronson, Arun, Asali, Ascenzi, Ashton, Aston, Astone, Aubin,
  Aufmuth, AultONeal, Austin, Avendano, Babak, Bacon, Badaracco, Bader, Bae,
  Baer, Baird, Baldaccini, Ballardin, Ballmer, Bals, Balsamo, Baltus, Banagiri,
  Bankar, Bankar, Barayoga, Barbieri, Barish, Barker, Barkett, Barneo, Barone,
  Barr, Barsotti, Barsuglia, Barta, Bartlett, Bartos, Bassiri, Basti, Bawaj,
  Bayley, Bazzan, B\'ecsy, Bejger, Belahcene, Bell, Beniwal, Benjamin, Bentley,
  Bergamin, Berger, Bergmann, Bernuzzi, Berry, Bersanetti, Bertolini,
  Betzwieser, Bhandare, Bhandari, Bidler, Biggs, Bilenko, Billingsley, Birney,
  Birnholtz, Biscans, Bischi, Biscoveanu, Bisht, Bissenbayeva, Bitossi,
  Bizouard, Blackburn, Blackman, Blair, Blair, Blair, Bobba, Bode, Boer,
  Boetzel, Bogaert, Bondu, Bonilla, Bonnand, Booker, Boom, Bork, Boschi, Bose,
  Bossilkov, Bosveld, Bouffanais, Bozzi, Bradaschia, Brady, Bramley, Branchesi,
  Brau, Breschi, Briant, Briggs, Brighenti, Brillet, Brinkmann, Brockill,
  Brooks, Brooks, Brown, Brunett, Bruno, Bruntz, Buikema, Bulik, Bulten,
  Buonanno, Buscicchio, Buskulic, Byer, Cabero, Cadonati, Cagnoli, Cahillane,
  Calder\'on~Bustillo, Callaghan, Callister, Calloni, Camp, Canepa, Cannon,
  Cao, Cao, Carapella, Carbognani, Caride, Carney, Carullo, Casanueva~Diaz,
  Casentini, Casta\~neda, Caudill, Cavagli\`a, Cavalier, Cavalieri, Cella,
  Cerd\'a-Dur\'an, Cesarini, Chaibi, Chakravarti, Chan, Chan, Chandra, Chao,
  Charlton, Chase, Chassande-Mottin, Chatterjee, Chaturvedi, Chatziioannou,
  Chen, Chen, Chen, Cheng, Cheong, Chia, Chiadini, Chierici, Chincarini,
  Chiummo, Cho, Cho, Cho, Christensen, Chu, Chua, Chung, Chung, Ciani,
  Ciecielag, Cie\ifmmode~\acute{s}\else \'{s}\fi{}lar, Ciobanu, Ciolfi,
  Cipriano, Cirone, Clara, Clark, Clearwater, Clesse, Cleva, Coccia, Cohadon,
  Cohen, Colleoni, Collette, Collins, Colpi, Constancio, Conti, Cooper, Corban,
  Corbitt, Cordero-Carri\'on, Corezzi, Corley, Cornish, Corre, Corsi, Cortese,
  Costa, Cotesta, Coughlin, Coughlin, Coulon, Countryman, Couvares, Covas,
  Coward, Cowart, Coyne, Coyne, Creighton, Creighton, Cripe, Croquette,
  Crowder, Cudell, Cullen, Cumming, Cummings, Cunningham, Cuoco, Curylo,
  Canton, D\'alya, Dana, Daneshgaran-Bajastani, D'Angelo, Danilishin,
  D'Antonio, Danzmann, Darsow-Fromm, Dasgupta, Datrier, Dattilo, Dave, Davier,
  Davies, Davis, Daw, DeBra, Deenadayalan, Degallaix, De~Laurentis,
  Del\'eglise, Delfavero, De~Lillo, Del~Pozzo, DeMarchi, D'Emilio, Demos, Dent,
  De~Pietri, De~Rosa, De~Rossi, DeSalvo, de~Varona, Dhurandhar, D\'{\i}az,
  Diaz-Ortiz, Dietrich, Di~Fiore, Di~Fronzo, Di~Giorgio, Di~Giovanni,
  Di~Giovanni, Di~Girolamo, Di~Lieto, Ding, Di~Pace, Di~Palma, Di~Renzo,
  Divakarla, Dmitriev, Doctor, Donovan, Dooley, Doravari, Dorrington, Downes,
  Drago, Driggers, Du, Ducoin, Dupej, Durante, D'Urso, Dwyer, Easter, Eddolls,
  Edelman, Edo, Edy, Effler, Ehrens, Eichholz, Eikenberry, Eisenmann,
  Eisenstein, Ejlli, Errico, Essick, Estelles, Estevez, Etienne, Etzel, Evans,
  Evans, Ewing, Fafone, Fairhurst, Fan, Farinon, Farr, Farr, Fauchon-Jones,
  Favata, Fays, Fazio, Feicht, Fejer, Feng, Fenyvesi, Ferguson,
  Fernandez-Galiana, Ferrante, Ferreira, Ferreira, Fidecaro, Fiori, Fiorucci,
  Fishbach, Fisher, Fittipaldi, Fitz-Axen, Fiumara, Flaminio, Floden, Flynn,
  Fong, Font, Forsyth, Fournier, Frasca, Frasconi, Frei, Freise, Frey, Frey,
  Fritschel, Frolov, Fronz\`e, Fulda, Fyffe, Gabbard, Gadre, Gaebel, Gair,
  Galaudage, Ganapathy, Ganguly, Gaonkar, Garc\'{\i}a-Quir\'os, Garufi,
  Gateley, Gaudio, Gayathri, Gemme, Genin, Gennai, George, George, Gergely,
  Ghonge, Ghosh, Ghosh, Ghosh, Giacomazzo, Giaime, Giardina, Gibson, Gier,
  Gill, Glanzer, Gniesmer, Godwin, Goetz, Goetz, Gohlke, Goncharov, Gonz\'alez,
  Gopakumar, Gossan, Gosselin, Gouaty, Grace, Grado, Granata, Grant, Gras,
  Grassia, Gray, Gray, Greco, Green, Green, Gretarsson, Griggs, Grignani,
  Grimaldi, Grimm, Grote, Grunewald, Gruning, Guidi, Guimaraes, Guix\'e,
  Gulati, Guo, Gupta, Gupta, Gupta, Gustafson, Gustafson, Haegel, Halim, Hall,
  Hamilton, Hammond, Haney, Hanke, Hanks, Hanna, Hannam, Hannuksela, Hansen,
  Hanson, Harder, Hardwick, Haris, Harms, Harry, Harry, Hasskew, Haster,
  Haughian, Hayes, Healy, Heidmann, Heintze, Heinze, Heitmann, Hellman, Hello,
  Hemming, Hendry, Heng, Hennes, Hennig, Heurs, Hild, Hinderer, Hoback,
  Hochheim, Hofgard, Hofman, Holgado, Holland, Holt, Holz, Hopkins, Horst,
  Hough, Howell, Hoy, Huang, H\"ubner, Huerta, Huet, Hughey, Hui, Husa,
  Huttner, Huxford, Huynh-Dinh, Idzkowski, Iess, Inchauspe, Ingram, Intini,
  Isac, Isi, Iyer, Jacqmin, Jadhav, Jadhav, James, Jani, Janthalur, Jaranowski,
  Jariwala, Jaume, Jenkins, Jiang, Johns, Johnson-McDaniel, Jones, Jones,
  Jones, Jones, Jones, Jonker, Ju, Junker, Kalaghatgi, Kalogera, Kamai,
  Kandhasamy, Kang, Kanner, Kapadia, Karki, Kashyap, Kasprzack, Kastaun,
  Katsanevas, Katsavounidis, Katzman, Kaufer, Kawabe, K\'ef\'elian, Keitel,
  Keivani, Kennedy, Key, Khadka, Khalili, Khan, Khan, Khan, Khazanov, Khetan,
  Khursheed, Kijbunchoo, Kim, Kim, Kim, Kim, Kim, Kim, Kim, Kimball, King,
  Kinley-Hanlon, Kirchhoff, Kissel, Kleybolte, Klimenko, Knowles, Knyazev,
  Koch, Koehlenbeck, Koekoek, Koley, Kondrashov, Kontos, Koper, Korobko, Korth,
  Kovalam, Kozak, Kringel, Krishnendu, Kr\'olak, Krupinski, Kuehn, Kumar,
  Kumar, Kumar, Kumar, Kumar, Kuo, Kutynia, Lackey, Laghi, Lalande, Lam,
  Lamberts, Landry, Lane, Lang, Lange, Lantz, Lanza, La~Rosa, Lartaux-Vollard,
  Lasky, Laxen, Lazzarini, Lazzaro, Leaci, Leavey, Lecoeuche, Lee, Lee, Lee,
  Lee, Lee, Lehmann, Leroy, Letendre, Levin, Li, Li, li, Li, Li, Linde, Linker,
  Linley, Littenberg, Liu, Liu, Llorens-Monteagudo, Lo, Lockwood, London,
  Longo, Lorenzini, Loriette, Lormand, Losurdo, Lough, Lousto, Lovelace,
  L\"uck, Lumaca, Lundgren, Ma, Macas, Macfoy, MacInnis, Macleod, MacMillan,
  Macquet, Maga\~na Hernandez, Maga\~na Sandoval, Magee, Majorana, Maksimovic,
  Malik, Man, Mandic, Mangano, Mansell, Manske, Mantovani, Mapelli, Marchesoni,
  Marion, M\'arka, M\'arka, Markakis, Markosyan, Markowitz, Maros, Marquina,
  Marsat, Martelli, Martin, Martin, Martinez, Martynov, Masalehdan, Mason,
  Massera, Masserot, Massinger, Masso-Reid, Mastrogiovanni, Matas, Matichard,
  Mavalvala, Maynard, McCann, McCarthy, McClelland, McCormick, McCuller,
  McGuire, McIsaac, McIver, McManus, McRae, McWilliams, Meacher, Meadors,
  Mehmet, Mehta, Mejuto~Villa, Melatos, Mendell, Mercer, Mereni, Merfeld,
  Merilh, Merritt, Merzougui, Meshkov, Messenger, Messick, Metzdorff, Meyers,
  Meylahn, Mhaske, Miani, Miao, Michaloliakos, Michel, Middleton, Milano,
  Miller, Millhouse, Mills, Milotti, Milovich-Goff, Minazzoli, Minenkov,
  Mishkin, Mishra, Mistry, Mitra, Mitrofanov, Mitselmakher, Mittleman, Mo,
  Mogushi, Mohapatra, Mohite, Molina-Ruiz, Mondin, Montani, Moore, Moraru,
  Morawski, Moreno, Morisaki, Mours, Mow-Lowry, Mozzon, Muciaccia, Mukherjee,
  Mukherjee, Mukherjee, Mukherjee, Mukund, Mullavey, Munch, Mu\~niz, Murray,
  Nagar, Nardecchia, Naticchioni, Nayak, Neil, Neilson, Nelemans, Nelson, Nery,
  Neunzert, Ng, Ng, Nguyen, Nguyen, Nichols, Nichols, Nissanke, Nitz, Nocera,
  Noh, North, Nothard, Nuttall, Oberling, O'Brien, Oganesyan, Ogin, Oh, Oh,
  Ohme, Ohta, Okada, Oliver, Olivetto, Oppermann, Oram, O'Reilly, Ormiston,
  Ortega, O'Shaughnessy, Ossokine, Osthelder, Ottaway, Overmier, Owen, Pace,
  Pagano, Page, Pagliaroli, Pai, Pai, Palamos, Palashov, Palomba, Pan, Panda,
  Pang, Pankow, Pannarale, Pant, Paoletti, Paoli, Parida, Parker, Pascucci,
  Pasqualetti, Passaquieti, Passuello, Patricelli, Payne, Pearlstone, Pechsiri,
  Pedersen, Pedraza, Pele, Penn, Perego, Perez, P\'erigois, Perreca, Perri\`es,
  Petermann, Pfeiffer, Phelps, Phukon, Piccinni, Pichot, Piendibene,
  Piergiovanni, Pierro, Pillant, Pinard, Pinto, Piotrzkowski, Pirello, Pitkin,
  Plastino, Poggiani, Pong, Ponrathnam, Popolizio, Porter, Powell, Prajapati,
  Prasai, Prasanna, Pratten, Prestegard, Principe, Prodi, Prokhorov, Punturo,
  Puppo, P\"urrer, Qi, Quetschke, Quinonez, Raab, Raaijmakers, Radkins,
  Radulesco, Raffai, Rafferty, Raja, Rajan, Rajbhandari, Rakhmanov, Ramirez,
  Ramos-Buades, Rana, Rao, Rapagnani, Raymond, Razzano, Read, Regimbau, Rei,
  Reid, Reitze, Rettegno, Ricci, Richardson, Richardson, Ricker,
  Riemenschneider, Riles, Rizzo, Robertson, Robinet, Rocchi, Rodriguez-Soto,
  Rolland, Rollins, Roma, Romanelli, Romano, Romel, Romero-Shaw, Romie, Rose,
  Rose, Rose, Rosi\ifmmode~\acute{n}\else \'{n}\fi{}ska, Rosofsky, Ross, Rowan,
  Rowlinson, Roy, Roy, Roy, Ruggi, Rutins, Ryan, Sachdev, Sadecki,
  Sakellariadou, Salafia, Salconi, Saleem, Salemi, Samajdar, Sanchez, Sanchez,
  Sanchis-Gual, Sanders, Santiago, Santos, Sarin, Sassolas, Sathyaprakash,
  Sauter, Savage, Savant, Sawant, Sayah, Schaetzl, Schale, Scheel, Scheuer,
  Schmidt, Schnabel, Schofield, Sch\"onbeck, Schreiber, Schulte, Schutz,
  Schwarm, Schwartz, Scott, Scott, Seidel, Sellers, Sengupta, Sennett,
  Sentenac, Sequino, Sergeev, Setyawati, Shaddock, Shaffer, Sharifi, Shahriar,
  Sharma, Sharma, Shawhan, Shen, Shikauchi, Shink, Shoemaker, Shoemaker,
  Shukla, ShyamSundar, Siellez, Sieniawska, Sigg, Singer, Singh, Singh, Singha,
  Singhal, Sintes, Sipala, Skliris, Slagmolen, Slaven-Blair, Smetana, Smith,
  Smith, Somala, Son, Soni, Sorazu, Sordini, Sorrentino, Souradeep, Sowell,
  Spencer, Spera, Srivastava, Srivastava, Staats, Stachie, Standke, Steer,
  Steinke, Steinlechner, Steinlechner, Steinmeyer, Stevenson, Stocks, Stops,
  Stover, Strain, Stratta, Strunk, Sturani, Stuver, Sudhagar, Sudhir,
  Summerscales, Sun, Sunil, Sur, Suresh, Sutton, Swinkels,
  Szczepa\ifmmode~\acute{n}\else \'{n}\fi{}czyk, Tacca, Tait, Talbot,
  Tanasijczuk, Tanner, Tao, T\'apai, Tapia, Tapia San~Martin, Tasson, Taylor,
  Tenorio, Terkowski, Thirugnanasambandam, Thomas, Thomas, Thompson, Thondapu,
  Thorne, Thrane, Tinsman, Saravanan, Tiwari, Tiwari, Tiwari, Toland, Tonelli,
  Tornasi, Torres-Forn\'e, Torrie, Tosta~e Melo, T\"oyr\"a, Travasso, Traylor,
  Tringali, Tripathee, Trovato, Trudeau, Tsang, Tse, Tso, Tsukada, Tsuna,
  Tsutsui, Turconi, Ubhi, Udall, Ueno, Ugolini, Unnikrishnan, Urban, Usman,
  Utina, Vahlbruch, Vajente, Valdes, Valentini, van Bakel, van Beuzekom,
  van~den Brand, Van Den~Broeck, Vander-Hyde, van~der Schaaf, Van~Heijningen,
  van Veggel, Vardaro, Varma, Vass, Vas\'uth, Vecchio, Vedovato, Veitch,
  Veitch, Venkateswara, Venugopalan, Verkindt, Veske, Vetrano, Vicer\'e, Viets,
  Vinciguerra, Vine, Vinet, Vitale, Vivanco, Vo, Vocca, Vorvick, Vyatchanin,
  Wade, Wade, Wade, Walet, Walker, Wallace, Wallace, Walsh, Wang, Wang, Wang,
  Ward, Warden, Warner, Was, Watchi, Weaver, Wei, Weinert, Weinstein, Weiss,
  Wellmann, Wen, We\ss{}els, Westhouse, Wette, Whelan, Whiting, Whittle,
  Wilken, Williams, Williamson, Willis, Willke, Winkler, Wipf, Wittel, Woan,
  Woehler, Wofford, Wong, Wright, Wu, Wysocki, Xiao, Yamamoto, Yang, Yang,
  Yang, Yap, Yazback, Yeeles, Yu, Yu, Yuen, Zadro\ifmmode~\dot{z}\else
  \.{z}\fi{}ny, Zadro\ifmmode~\dot{z}\else \.{z}\fi{}ny, Zanolin, Zelenova,
  Zendri, Zevin, Zhang, Zhang, Zhang, Zhao, Zhao, Zhou, Zhou, Zhu, Zimmerman,
  Zucker, \& Zweizig}]{Abbott+20}
Abbott, R., Abbott, T.~D., Abraham, S., {et~al.} 2020, Phys. Rev. Lett., 125,
  101102, \dodoi{10.1103/PhysRevLett.125.101102}

\bibitem[{{Aranzana} {et~al.}(2018){Aranzana}, {K{\"o}rding}, {Uttley},
  {Scaringi}, \& {Bloemen}}]{Aranzana+18}
{Aranzana}, E., {K{\"o}rding}, E., {Uttley}, P., {Scaringi}, S., \& {Bloemen},
  S. 2018, \mnras, 476, 2501, \dodoi{10.1093/mnras/sty413}

\bibitem[{{Astropy Collaboration} {et~al.}(2013){Astropy Collaboration},
  {Robitaille}, {Tollerud}, {Greenfield}, {Droettboom}, {Bray}, {Aldcroft},
  {Davis}, {Ginsburg}, {Price-Whelan}, {Kerzendorf}, {Conley}, {Crighton},
  {Barbary}, {Muna}, {Ferguson}, {Grollier}, {Parikh}, {Nair}, {Unther},
  {Deil}, {Woillez}, {Conseil}, {Kramer}, {Turner}, {Singer}, {Fox}, {Weaver},
  {Zabalza}, {Edwards}, {Azalee Bostroem}, {Burke}, {Casey}, {Crawford},
  {Dencheva}, {Ely}, {Jenness}, {Labrie}, {Lim}, {Pierfederici}, {Pontzen},
  {Ptak}, {Refsdal}, {Servillat}, \& {Streicher}}]{Astropy13}
{Astropy Collaboration}, {Robitaille}, T.~P., {Tollerud}, E.~J., {et~al.} 2013,
  \aap, 558, A33, \dodoi{10.1051/0004-6361/201322068}

\bibitem[{{Astropy Collaboration} {et~al.}(2018){Astropy Collaboration},
  {Price-Whelan}, {Sip{\H{o}}cz}, {G{\"u}nther}, {Lim}, {Crawford}, {Conseil},
  {Shupe}, {Craig}, {Dencheva}, {Ginsburg}, {VanderPlas}, {Bradley},
  {P{\'e}rez-Su{\'a}rez}, {de Val-Borro}, {Aldcroft}, {Cruz}, {Robitaille},
  {Tollerud}, {Ardelean}, {Babej}, {Bach}, {Bachetti}, {Bakanov}, {Bamford},
  {Barentsen}, {Barmby}, {Baumbach}, {Berry}, {Biscani}, {Boquien}, {Bostroem},
  {Bouma}, {Brammer}, {Bray}, {Breytenbach}, {Buddelmeijer}, {Burke},
  {Calderone}, {Cano Rodr{\'\i}guez}, {Cara}, {Cardoso}, {Cheedella}, {Copin},
  {Corrales}, {Crichton}, {D'Avella}, {Deil}, {Depagne}, {Dietrich}, {Donath},
  {Droettboom}, {Earl}, {Erben}, {Fabbro}, {Ferreira}, {Finethy}, {Fox},
  {Garrison}, {Gibbons}, {Goldstein}, {Gommers}, {Greco}, {Greenfield},
  {Groener}, {Grollier}, {Hagen}, {Hirst}, {Homeier}, {Horton}, {Hosseinzadeh},
  {Hu}, {Hunkeler}, {Ivezi{\'c}}, {Jain}, {Jenness}, {Kanarek}, {Kendrew},
  {Kern}, {Kerzendorf}, {Khvalko}, {King}, {Kirkby}, {Kulkarni}, {Kumar},
  {Lee}, {Lenz}, {Littlefair}, {Ma}, {Macleod}, {Mastropietro}, {McCully},
  {Montagnac}, {Morris}, {Mueller}, {Mumford}, {Muna}, {Murphy}, {Nelson},
  {Nguyen}, {Ninan}, {N{\"o}the}, {Ogaz}, {Oh}, {Parejko}, {Parley}, {Pascual},
  {Patil}, {Patil}, {Plunkett}, {Prochaska}, {Rastogi}, {Reddy Janga},
  {Sabater}, {Sakurikar}, {Seifert}, {Sherbert}, {Sherwood-Taylor}, {Shih},
  {Sick}, {Silbiger}, {Singanamalla}, {Singer}, {Sladen}, {Sooley},
  {Sornarajah}, {Streicher}, {Teuben}, {Thomas}, {Tremblay}, {Turner},
  {Terr{\'o}n}, {van Kerkwijk}, {de la Vega}, {Watkins}, {Weaver}, {Whitmore},
  {Woillez}, {Zabalza}, \& {Astropy Contributors}}]{Astropy18}
{Astropy Collaboration}, {Price-Whelan}, A.~M., {Sip{\H{o}}cz}, B.~M., {et~al.}
  2018, \aj, 156, 123, \dodoi{10.3847/1538-3881/aabc4f}

\bibitem[{{Ba{\~n}ados} {et~al.}(2018){Ba{\~n}ados}, {Venemans},
  {Mazzucchelli}, {Farina}, {Walter}, {Wang}, {Decarli}, {Stern}, {Fan},
  {Davies}, {Hennawi}, {Simcoe}, {Turner}, {Rix}, {Yang}, {Kelson}, {Rudie}, \&
  {Winters}}]{Banados+18}
{Ba{\~n}ados}, E., {Venemans}, B.~P., {Mazzucchelli}, C., {et~al.} 2018, \nat,
  553, 473, \dodoi{10.1038/nature25180}

\bibitem[{{Bae} \& {Woo}(2014)}]{Bae&Woo14}
{Bae}, H.-J., \& {Woo}, J.-H. 2014, \apj, 795, 30,
  \dodoi{10.1088/0004-637X/795/1/30}

\bibitem[{{Baldassare} {et~al.}(2018){Baldassare}, {Geha}, \&
  {Greene}}]{Baldassare+18}
{Baldassare}, V.~F., {Geha}, M., \& {Greene}, J. 2018, \apj, 868, 152,
  \dodoi{10.3847/1538-4357/aae6cf}

\bibitem[{{Baldassare} {et~al.}(2015){Baldassare}, {Reines}, {Gallo}, \&
  {Greene}}]{Baldassare+15}
{Baldassare}, V.~F., {Reines}, A.~E., {Gallo}, E., \& {Greene}, J.~E. 2015,
  \apjl, 809, L14, \dodoi{10.1088/2041-8205/809/1/L14}

\bibitem[{{Baumgardt}(2017)}]{Baumgardt17}
{Baumgardt}, H. 2017, \mnras, 464, 2174, \dodoi{10.1093/mnras/stw2488}

\bibitem[{{Bentz} {et~al.}(2013){Bentz}, {Denney}, {Grier}, {Barth},
  {Peterson}, {Vestergaard}, {Bennert}, {Canalizo}, {De Rosa}, {Filippenko},
  {Gates}, {Greene}, {Li}, {Malkan}, {Pogge}, {Stern}, {Treu}, \&
  {Woo}}]{Bentz+13}
{Bentz}, M.~C., {Denney}, K.~D., {Grier}, C.~J., {et~al.} 2013, \apj, 767, 149,
  \dodoi{10.1088/0004-637X/767/2/149}

\bibitem[{{Bertin} \& {Arnouts}(1996)}]{Bertin&Arnouts96}
{Bertin}, E., \& {Arnouts}, S. 1996, \aaps, 117, 393,
  \dodoi{10.1051/aas:1996164}

\bibitem[{{Bertin} {et~al.}(2002){Bertin}, {Mellier}, {Radovich}, {Missonnier},
  {Didelon}, \& {Morin}}]{Bertin+02}
{Bertin}, E., {Mellier}, Y., {Radovich}, M., {et~al.} 2002, in Astronomical
  Society of the Pacific Conference Series, Vol. 281, Astronomical Data
  Analysis Software and Systems XI, ed. D.~A. {Bohlender}, D.~{Durand}, \&
  T.~H. {Handley}, 228

\bibitem[{{Cappellari} \& {Emsellem}(2004)}]{Cappellari&Emsellem04}
{Cappellari}, M., \& {Emsellem}, E. 2004, \pasp, 116, 138,
  \dodoi{10.1086/381875}

\bibitem[{{Chilingarian} {et~al.}(2018){Chilingarian}, {Katkov}, {Zolotukhin},
  {Grishin}, {Beletsky}, {Boutsia}, \& {Osip}}]{Chilingarian+18}
{Chilingarian}, I.~V., {Katkov}, I.~Y., {Zolotukhin}, I.~Y., {et~al.} 2018,
  \apj, 863, 1, \dodoi{10.3847/1538-4357/aad184}

\bibitem[{{Chilingarian} {et~al.}(2017){Chilingarian}, {Zolotukhin}, {Katkov},
  {Melchior}, {Rubtsov}, \& {Grishin}}]{Chilingarian+17}
{Chilingarian}, I.~V., {Zolotukhin}, I.~Y., {Katkov}, I.~Y., {et~al.} 2017,
  \apjs, 228, 14, \dodoi{10.3847/1538-4365/228/2/14}

\bibitem[{{Cseh} {et~al.}(2015){Cseh}, {Webb}, {Godet}, {Barret}, {Corbel},
  {Coriat}, {Falcke}, {Farrell}, {K{\"o}rding}, {Lenc}, \& {Wrobel}}]{Cseh+15}
{Cseh}, D., {Webb}, N.~A., {Godet}, O., {et~al.} 2015, \mnras, 446, 3268,
  \dodoi{10.1093/mnras/stu2363}

\bibitem[{{Dong} {et~al.}(2012){Dong}, {Ho}, {Yuan}, {Wang}, {Fan}, {Zhou}, \&
  {Jiang}}]{Dong+12}
{Dong}, X.-B., {Ho}, L.~C., {Yuan}, W., {et~al.} 2012, \apj, 755, 167,
  \dodoi{10.1088/0004-637X/755/2/167}

\bibitem[{{Elvis} {et~al.}(1978){Elvis}, {Maccacaro}, {Wilson}, {Ward},
  {Penston}, {Fosbury}, \& {Perola}}]{Elvis+78}
{Elvis}, M., {Maccacaro}, T., {Wilson}, A.~S., {et~al.} 1978, \mnras, 183, 129,
  \dodoi{10.1093/mnras/183.2.129}

\bibitem[{{Eun} {et~al.}(2017){Eun}, {Woo}, \& {Bae}}]{Eun+17}
{Eun}, D.-i., {Woo}, J.-H., \& {Bae}, H.-J. 2017, \apj, 842, 5,
  \dodoi{10.3847/1538-4357/aa6daf}

\bibitem[{{Filippenko} \& {Ho}(2003)}]{Filippenko&Ho03}
{Filippenko}, A.~V., \& {Ho}, L.~C. 2003, \apjl, 588, L13,
  \dodoi{10.1086/375361}

\bibitem[{{Gebhardt} {et~al.}(2000){Gebhardt}, {Pryor}, {O'Connell},
  {Williams}, \& {Hesser}}]{Gebhardt+00}
{Gebhardt}, K., {Pryor}, C., {O'Connell}, R.~D., {Williams}, T.~B., \&
  {Hesser}, J.~E. 2000, \aj, 119, 1268, \dodoi{10.1086/301275}

\bibitem[{{Gerssen} {et~al.}(2002){Gerssen}, {van der Marel}, {Gebhardt},
  {Guhathakurta}, {Peterson}, \& {Pryor}}]{Gerssen+02}
{Gerssen}, J., {van der Marel}, R.~P., {Gebhardt}, K., {et~al.} 2002, \aj, 124,
  3270, \dodoi{10.1086/344584}

\bibitem[{{Graur} \& {Maoz}(2013)}]{Graur&Maoz13}
{Graur}, O., \& {Maoz}, D. 2013, \mnras, 430, 1746,
  \dodoi{10.1093/mnras/sts718}

\bibitem[{{Greene}(2012)}]{Greene12}
{Greene}, J.~E. 2012, Nature Communications, 3, 1304,
  \dodoi{10.1038/ncomms2314}

\bibitem[{{Greene} \& {Ho}(2004)}]{Greene&Ho04}
{Greene}, J.~E., \& {Ho}, L.~C. 2004, \apj, 610, 722, \dodoi{10.1086/421719}

\bibitem[{{Greene} \& {Ho}(2005)}]{Greene&ho05}
---. 2005, \apj, 630, 122, \dodoi{10.1086/431897}

\bibitem[{{Greene} \& {Ho}(2007)}]{Greene&Ho07}
---. 2007, \apj, 670, 92, \dodoi{10.1086/522082}

\bibitem[{{Greene} {et~al.}(2020){Greene}, {Strader}, \& {Ho}}]{Greene+20}
{Greene}, J.~E., {Strader}, J., \& {Ho}, L.~C. 2020, \araa, 58, 257,
  \dodoi{10.1146/annurev-astro-032620-021835}

\bibitem[{{Izotov} \& {Thuan}(2007)}]{Izotov&Thuan07}
{Izotov}, Y.~I., \& {Thuan}, T.~X. 2007, \apj, 665, 1115,
  \dodoi{10.1086/519922}

\bibitem[{{Izotov} \& {Thuan}(2009{\natexlab{a}})}]{Izotov&Thuan09a}
---. 2009{\natexlab{a}}, \apj, 690, 1797, \dodoi{10.1088/0004-637X/690/2/1797}

\bibitem[{{Izotov} \& {Thuan}(2009{\natexlab{b}})}]{Izotov&Thuan09b}
---. 2009{\natexlab{b}}, \apj, 707, 1560, \dodoi{10.1088/0004-637X/707/2/1560}

\bibitem[{{Izotov} {et~al.}(2007){Izotov}, {Thuan}, \& {Guseva}}]{Izotov+07}
{Izotov}, Y.~I., {Thuan}, T.~X., \& {Guseva}, N.~G. 2007, \apj, 671, 1297,
  \dodoi{10.1086/522923}

\bibitem[{{Kass} \& {Raftery}(1995)}]{Kass&Raftery95}
{Kass}, R.~E., \& {Raftery}, A.~E. 1995, J. Am. Stat. Assoc., 90, 773,
  \dodoi{10.2307/2291091}

\bibitem[{{Kauffmann} {et~al.}(2003){Kauffmann}, {Heckman}, {Tremonti},
  {Brinchmann}, {Charlot}, {White}, {Ridgway}, {Brinkmann}, {Fukugita}, {Hall},
  {Ivezi{\'c}}, {Richards}, \& {Schneider}}]{Kauffmann+03}
{Kauffmann}, G., {Heckman}, T.~M., {Tremonti}, C., {et~al.} 2003, \mnras, 346,
  1055, \dodoi{10.1111/j.1365-2966.2003.07154.x}

\bibitem[{{Kim} {et~al.}(2018){Kim}, {Karouzos}, {Im}, {Choi}, {Kim}, {Jun},
  {Lee}, \& {Mezcua}}]{Kim+18}
{Kim}, J., {Karouzos}, M., {Im}, M., {et~al.} 2018, Journal of Korean
  Astronomical Society, 51, 89, \dodoi{10.5303/JKAS.2018.51.4.89}

\bibitem[{{Kormendy} \& {Ho}(2013)}]{Kormendy&Ho13}
{Kormendy}, J., \& {Ho}, L.~C. 2013, \araa, 51, 511,
  \dodoi{10.1146/annurev-astro-082708-101811}

\bibitem[{{Kormendy} \& {Richstone}(1995)}]{Kormendy&Richstone95}
{Kormendy}, J., \& {Richstone}, D. 1995, \araa, 33, 581,
  \dodoi{10.1146/annurev.aa.33.090195.003053}

\bibitem[{{Lang} {et~al.}(2010){Lang}, {Hogg}, {Mierle}, {Blanton}, \&
  {Roweis}}]{Lang+10}
{Lang}, D., {Hogg}, D.~W., {Mierle}, K., {Blanton}, M., \& {Roweis}, S. 2010,
  \aj, 139, 1782, \dodoi{10.1088/0004-6256/139/5/1782}

\bibitem[{{Latif} {et~al.}(2013){Latif}, {Schleicher}, {Schmidt}, \&
  {Niemeyer}}]{Latif+13}
{Latif}, M.~A., {Schleicher}, D.~R.~G., {Schmidt}, W., \& {Niemeyer}, J.~C.
  2013, \mnras, 436, 2989, \dodoi{10.1093/mnras/stt1786}

\bibitem[{{Le} {et~al.}(2020){Le}, {Woo}, \& {Xue}}]{Le+20}
{Le}, H. A.~N., {Woo}, J.-H., \& {Xue}, Y. 2020, \apj, 901, 35,
  \dodoi{10.3847/1538-4357/abada0}

\bibitem[{{Liu} {et~al.}(2018){Liu}, {Yuan}, {Dong}, {Zhou}, \& {Liu}}]{Liu+18}
{Liu}, H.-Y., {Yuan}, W., {Dong}, X.-B., {Zhou}, H., \& {Liu}, W.-J. 2018,
  \apjs, 235, 40, \dodoi{10.3847/1538-4365/aab88e}

\bibitem[{{Loeb} \& {Rasio}(1994)}]{Loeb&Rasio94}
{Loeb}, A., \& {Rasio}, F.~A. 1994, \apj, 432, 52, \dodoi{10.1086/174548}

\bibitem[{{Mart{\'\i}nez-Palomera} {et~al.}(2020){Mart{\'\i}nez-Palomera},
  {Lira}, {Bhalla-Ladd}, {F{\"o}rster}, \& {Plotkin}}]{Martinez+20}
{Mart{\'\i}nez-Palomera}, J., {Lira}, P., {Bhalla-Ladd}, I., {F{\"o}rster}, F.,
  \& {Plotkin}, R.~M. 2020, \apj, 889, 113, \dodoi{10.3847/1538-4357/ab5f5b}

\bibitem[{{McConnell} \& {Ma}(2013)}]{McConnell&Ma13}
{McConnell}, N.~J., \& {Ma}, C.-P. 2013, \apj, 764, 184,
  \dodoi{10.1088/0004-637X/764/2/184}

\bibitem[{{Mezcua}(2017)}]{Mezcua17}
{Mezcua}, M. 2017, International Journal of Modern Physics D, 26, 1730021,
  \dodoi{10.1142/S021827181730021X}

\bibitem[{{Miller} {et~al.}(2003){Miller}, {Fabbiano}, {Miller}, \&
  {Fabian}}]{Miller+03}
{Miller}, J.~M., {Fabbiano}, G., {Miller}, M.~C., \& {Fabian}, A.~C. 2003,
  \apjl, 585, L37, \dodoi{10.1086/368373}

\bibitem[{{Mortlock} {et~al.}(2011){Mortlock}, {Warren}, {Venemans}, {Patel},
  {Hewett}, {McMahon}, {Simpson}, {Theuns}, {Gonz{\'a}les-Solares}, {Adamson},
  {Dye}, {Hambly}, {Hirst}, {Irwin}, {Kuiper}, {Lawrence}, \&
  {R{\"o}ttgering}}]{Mortlock+11}
{Mortlock}, D.~J., {Warren}, S.~J., {Venemans}, B.~P., {et~al.} 2011, \nat,
  474, 616, \dodoi{10.1038/nature10159}

\bibitem[{{Mushotzky} {et~al.}(2011){Mushotzky}, {Edelson}, {Baumgartner}, \&
  {Gandhi}}]{Mushotzky+11}
{Mushotzky}, R.~F., {Edelson}, R., {Baumgartner}, W., \& {Gandhi}, P. 2011,
  \apjl, 743, L12, \dodoi{10.1088/2041-8205/743/1/L12}

\bibitem[{{Nucita} {et~al.}(2017){Nucita}, {Manni}, {De Paolis}, {Giordano}, \&
  {Ingrosso}}]{Nucita+17}
{Nucita}, A.~A., {Manni}, L., {De Paolis}, F., {Giordano}, M., \& {Ingrosso},
  G. 2017, \apj, 837, 66, \dodoi{10.3847/1538-4357/aa5f4f}

\bibitem[{{Onoue} {et~al.}(2019){Onoue}, {Kashikawa}, {Matsuoka}, {Kato},
  {Izumi}, {Nagao}, {Strauss}, {Harikane}, {Imanishi}, {Ito}, {Iwasawa},
  {Kawaguchi}, {Lee}, {Noboriguchi}, {Suh}, {Tanaka}, \& {Toba}}]{Onoue+19}
{Onoue}, M., {Kashikawa}, N., {Matsuoka}, Y., {et~al.} 2019, \apj, 880, 77,
  \dodoi{10.3847/1538-4357/ab29e9}

\bibitem[{{Park} {et~al.}(2012){Park}, {Woo}, {Treu}, {Barth}, {Bentz},
  {Bennert}, {Canalizo}, {Filippenko}, {Gates}, {Greene}, {Malkan}, \&
  {Walsh}}]{Park+12}
{Park}, D., {Woo}, J.-H., {Treu}, T., {et~al.} 2012, \apj, 747, 30,
  \dodoi{10.1088/0004-637X/747/1/30}

\bibitem[{{Peterson}(1993)}]{Peterson93}
{Peterson}, B.~M. 1993, \pasp, 105, 247, \dodoi{10.1086/133140}

\bibitem[{{Peterson} {et~al.}(2004){Peterson}, {Ferrarese}, {Gilbert}, {Kaspi},
  {Malkan}, {Maoz}, {Merritt}, {Netzer}, {Onken}, {Pogge}, {Vestergaard}, \&
  {Wandel}}]{Peterson+04}
{Peterson}, B.~M., {Ferrarese}, L., {Gilbert}, K.~M., {et~al.} 2004, \apj, 613,
  682, \dodoi{10.1086/423269}

\bibitem[{{Pooley} \& {Rappaport}(2006)}]{Pooley&Rappaport06}
{Pooley}, D., \& {Rappaport}, S. 2006, \apjl, 644, L45, \dodoi{10.1086/505344}

\bibitem[{{Rakshit} \& {Stalin}(2017)}]{Rakshit&Stalin17}
{Rakshit}, S., \& {Stalin}, C.~S. 2017, \apj, 842, 96,
  \dodoi{10.3847/1538-4357/aa72f4}

\bibitem[{{Rakshit} {et~al.}(2019){Rakshit}, {Woo}, {Gallo}, {Hodges-Kluck},
  {Shin}, {Jeon}, {Bae}, {Baldassare}, {Cho}, {Cho}, {Foord}, {Kang}, {Kang},
  {Karouzos}, {Kim}, {Kim}, {Le}, {Park}, {Park}, {Son}, {Sung}, {Bennert}, \&
  {Malkan}}]{Rakshit+19}
{Rakshit}, S., {Woo}, J.-H., {Gallo}, E., {et~al.} 2019, \apj, 886, 93,
  \dodoi{10.3847/1538-4357/ab49fd}

\bibitem[{{Reines} \& {Comastri}(2016)}]{Reines&Comastri16}
{Reines}, A.~E., \& {Comastri}, A. 2016, \pasa, 33, e054,
  \dodoi{10.1017/pasa.2016.46}

\bibitem[{{Reines} {et~al.}(2013){Reines}, {Greene}, \& {Geha}}]{Reines+13}
{Reines}, A.~E., {Greene}, J.~E., \& {Geha}, M. 2013, \apj, 775, 116,
  \dodoi{10.1088/0004-637X/775/2/116}

\bibitem[{{Remillard} \& {McClintock}(2006)}]{Remillard&McClintock06}
{Remillard}, R.~A., \& {McClintock}, J.~E. 2006, \araa, 44, 49,
  \dodoi{10.1146/annurev.astro.44.051905.092532}

\bibitem[{{Rodr{\'\i}guez-Pascual} {et~al.}(1997){Rodr{\'\i}guez-Pascual},
  {Alloin}, {Clavel}, {Crenshaw}, {Horne}, {Kriss}, {Krolik}, {Malkan},
  {Netzer}, {O'Brien}, {Peterson}, {Reichert}, {Wamsteker}, {Alexander},
  {Barr}, {Blandford}, {Bregman}, {Carone}, {Clements}, {Courvoisier}, {De
  Robertis}, {Dietrich}, {Dottori}, {Edelson}, {Filippenko}, {Gaskell},
  {Huchra}, {Hutchings}, {Kollatschny}, {Koratkar}, {Korista}, {Laor},
  {MacAlpine}, {Martin}, {Maoz}, {McCollum}, {Morris}, {Perola}, {Pogge},
  {Ptak}, {Recondo-Gonz{\'a}lez}, {Rodr{\'\i}guez-Espinoza}, {Rokaki},
  {Santos-Lle{\'o}}, {Sekiguchi}, {Shull}, {Snijders}, {Sparke}, {Stirpe},
  {Stoner}, {Sun}, {Wagner}, {Wanders}, {Wilkes}, {Winge}, \&
  {Zheng}}]{Rodriguez-Pascual+97}
{Rodr{\'\i}guez-Pascual}, P.~M., {Alloin}, D., {Clavel}, J., {et~al.} 1997,
  \apjs, 110, 9, \dodoi{10.1086/312996}

\bibitem[{{S{\'a}nchez-Bl{\'a}zquez} {et~al.}(2006){S{\'a}nchez-Bl{\'a}zquez},
  {Peletier}, {Jim{\'e}nez-Vicente}, {Cardiel}, {Cenarro},
  {Falc{\'o}n-Barroso}, {Gorgas}, {Selam}, \& {Vazdekis}}]{Sanchez-Blazquez+06}
{S{\'a}nchez-Bl{\'a}zquez}, P., {Peletier}, R.~F., {Jim{\'e}nez-Vicente}, J.,
  {et~al.} 2006, \mnras, 371, 703, \dodoi{10.1111/j.1365-2966.2006.10699.x}

\bibitem[{{Shen} \& {Liu}(2012)}]{Shen&liu12}
{Shen}, Y., \& {Liu}, X. 2012, \apj, 753, 125,
  \dodoi{10.1088/0004-637X/753/2/125}

\bibitem[{{Silk} \& {Arons}(1975)}]{Silk&Arons75}
{Silk}, J., \& {Arons}, J. 1975, \apjl, 200, L131, \dodoi{10.1086/181914}

\bibitem[{{Sutton} {et~al.}(2012){Sutton}, {Roberts}, {Walton}, {Gladstone}, \&
  {Scott}}]{Sutton+12}
{Sutton}, A.~D., {Roberts}, T.~P., {Walton}, D.~J., {Gladstone}, J.~C., \&
  {Scott}, A.~E. 2012, \mnras, 423, 1154,
  \dodoi{10.1111/j.1365-2966.2012.20944.x}

\bibitem[{{Thompson} {et~al.}(2019){Thompson}, {Kochanek}, {Stanek}, {Badenes},
  {Post}, {Jayasinghe}, {Latham}, {Bieryla}, {Esquerdo}, {Berlind}, {Calkins},
  {Tayar}, {Lindegren}, {Johnson}, {Holoien}, {Auchettl}, \&
  {Covey}}]{Thompson+19}
{Thompson}, T.~A., {Kochanek}, C.~S., {Stanek}, K.~Z., {et~al.} 2019, Science,
  366, 637, \dodoi{10.1126/science.aau4005}

\bibitem[{{Tody}(1986)}]{Tody86}
{Tody}, D. 1986, in Society of Photo-Optical Instrumentation Engineers (SPIE)
  Conference Series, Vol. 627, Instrumentation in astronomy VI, ed. D.~L.
  {Crawford}, 733, \dodoi{10.1117/12.968154}

\bibitem[{{Tremonti} {et~al.}(2004){Tremonti}, {Heckman}, {Kauffmann},
  {Brinchmann}, {Charlot}, {White}, {Seibert}, {Peng}, {Schlegel}, {Uomoto},
  {Fukugita}, \& {Brinkmann}}]{Tremonti+04}
{Tremonti}, C.~A., {Heckman}, T.~M., {Kauffmann}, G., {et~al.} 2004, \apj, 613,
  898, \dodoi{10.1086/423264}

\bibitem[{{van Dokkum}(2001)}]{vanDokkum01}
{van Dokkum}, P.~G. 2001, \pasp, 113, 1420, \dodoi{10.1086/323894}

\bibitem[{{Vanden Berk} {et~al.}(2006){Vanden Berk}, {Shen}, {Yip},
  {Schneider}, {Connolly}, {Burton}, {Jester}, {Hall}, {Szalay}, \&
  {Brinkmann}}]{VandenBerk+06}
{Vanden Berk}, D.~E., {Shen}, J., {Yip}, C.-W., {et~al.} 2006, \aj, 131, 84,
  \dodoi{10.1086/497973}

\bibitem[{{Virtanen} {et~al.}(2020){Virtanen}, {Gommers}, {Oliphant},
  {Haberland}, {Reddy}, {Cournapeau}, {Burovski}, {Peterson}, {Weckesser},
  {Bright}, {van der Walt}, {Brett}, {Wilson}, {Millman}, {Mayorov}, {Nelson},
  {Jones}, {Kern}, {Larson}, {Carey}, {Polat}, {Feng}, {Moore}, {VanderPlas},
  {Laxalde}, {Perktold}, {Cimrman}, {Henriksen}, {Quintero}, {Harris},
  {Archibald}, {Ribeiro}, {Pedregosa}, {van Mulbregt}, \& {SciPy 1. 0
  Contributors}}]{Virtanen+20}
{Virtanen}, P., {Gommers}, R., {Oliphant}, T.~E., {et~al.} 2020, Nature
  Methods, 17, 261, \dodoi{10.1038/s41592-019-0686-2}

\bibitem[{{Volonteri} {et~al.}(2008){Volonteri}, {Lodato}, \&
  {Natarajan}}]{Volonteri+08}
{Volonteri}, M., {Lodato}, G., \& {Natarajan}, P. 2008, \mnras, 383, 1079,
  \dodoi{10.1111/j.1365-2966.2007.12589.x}

\bibitem[{{Walsh} {et~al.}(2009){Walsh}, {Minezaki}, {Bentz}, {Barth},
  {Baliber}, {Li}, {Stern}, {Bennert}, {Brown}, {Canalizo}, {Filippenko},
  {Gates}, {Greene}, {Malkan}, {Sakata}, {Street}, {Treu}, {Woo}, \&
  {Yoshii}}]{Walsh+09}
{Walsh}, J.~L., {Minezaki}, T., {Bentz}, M.~C., {et~al.} 2009, \apjs, 185, 156,
  \dodoi{10.1088/0067-0049/185/1/156}

\bibitem[{{Webb} {et~al.}(2012){Webb}, {Cseh}, {Lenc}, {Godet}, {Barret},
  {Corbel}, {Farrell}, {Fender}, {Gehrels}, \& {Heywood}}]{Webb+12}
{Webb}, N., {Cseh}, D., {Lenc}, E., {et~al.} 2012, Science, 337, 554,
  \dodoi{10.1126/science.1222779}

\bibitem[{{Woo} {et~al.}(2016){Woo}, {Bae}, {Son}, \& {Karouzos}}]{Woo+16}
{Woo}, J.-H., {Bae}, H.-J., {Son}, D., \& {Karouzos}, M. 2016, \apj, 817, 108,
  \dodoi{10.3847/0004-637X/817/2/108}

\bibitem[{{Woo} {et~al.}(2019{\natexlab{a}}){Woo}, {Cho}, {Gallo},
  {Hodges-Kluck}, {Le}, {Shin}, {Son}, \& {Horst}}]{Woo+19a}
{Woo}, J.-H., {Cho}, H., {Gallo}, E., {et~al.} 2019{\natexlab{a}}, Nature
  Astronomy, 3, 755, \dodoi{10.1038/s41550-019-0790-3}

\bibitem[{{Woo} {et~al.}(2014){Woo}, {Kim}, {Park}, {Bae}, {Kim}, {Lee}, {Kim},
  \& {Kwon}}]{Woo+14}
{Woo}, J.-H., {Kim}, J.-G., {Park}, D., {et~al.} 2014, Journal of Korean
  Astronomical Society, 47, 167, \dodoi{10.5303/JKAS.2014.47.5.167}

\bibitem[{{Woo} \& {Urry}(2002)}]{Woo&Urry02}
{Woo}, J.-H., \& {Urry}, C.~M. 2002, \apj, 579, 530, \dodoi{10.1086/342878}

\bibitem[{{Woo} {et~al.}(2015){Woo}, {Yoon}, {Park}, {Park}, \& {Kim}}]{Woo+15}
{Woo}, J.-H., {Yoon}, Y., {Park}, S., {Park}, D., \& {Kim}, S.~C. 2015, \apj,
  801, 38, \dodoi{10.1088/0004-637X/801/1/38}

\bibitem[{{Woo} {et~al.}(2019{\natexlab{b}}){Woo}, {Son}, {Gallo},
  {Hodges-Kluck}, {Jeon}, {Shin}, {Bae}, {Cho}, {Cho}, {Kang}, {Kang},
  {Karouzos}, {Kim}, {Kim}, {Le}, {Park}, {Park}, {Rakshit}, \&
  {Sung}}]{Woo+19b}
{Woo}, J.-H., {Son}, D., {Gallo}, E., {et~al.} 2019{\natexlab{b}}, Journal of
  Korean Astronomical Society, 52, 109, \dodoi{10.5303/JKAS.2019.52.4.109}

\end{thebibliography}
\bibliographystyle{aasjournal}

\section*{Appendix \label{sec:app}}
We present the spectral decomposition results of the 24 broad H$\alpha$ candidates with M$_{\rm BH}$ $<$ $10^6$ \msun\ in Figure \ref{fig:fitting_1_8}. The result of candidate 1 is shown in Figure \ref{fig:fitting_25}. As explain in Section \ref{sec:spec}, IDs 2, 12, 14, 16, 20, 23, and 24 show the case that the model without H$\alpha$ broad component (second and forth column) cannot fit the H$\alpha$ line well and large residual is present between each emission line. The model without H$\alpha$ broad component for IDs 3, 5, and 17 is not able to fit the both H$\alpha$ and [S\RNum{2}] lines, and the model for the other candidates cannot fit the [S\RNum{2}] lines appropriately. They fit the [S\RNum{2}] lines much broader than the real, so a significant residual is shown between the two [S\RNum{2}] lines. The amplitude of H$\alpha$ broad component of the IDs 6, 12, 14, and 20 seems low relative to the amplitude of the emission lines. However their absolute amplitude is larger than their flux density error and the residual around the emission lines, so adding the H$\alpha$ broad component is reliable. Also note that, according the criteria of Section \ref{sec:spec}, the BIC and $\chi^2_{red}$ of their fitting results decrease by 10 when the broad component is added, and the A/N of each broad component is larger than 10.

\begin{figure*}
    \centering
    \includegraphics[angle=0,width=175mm]{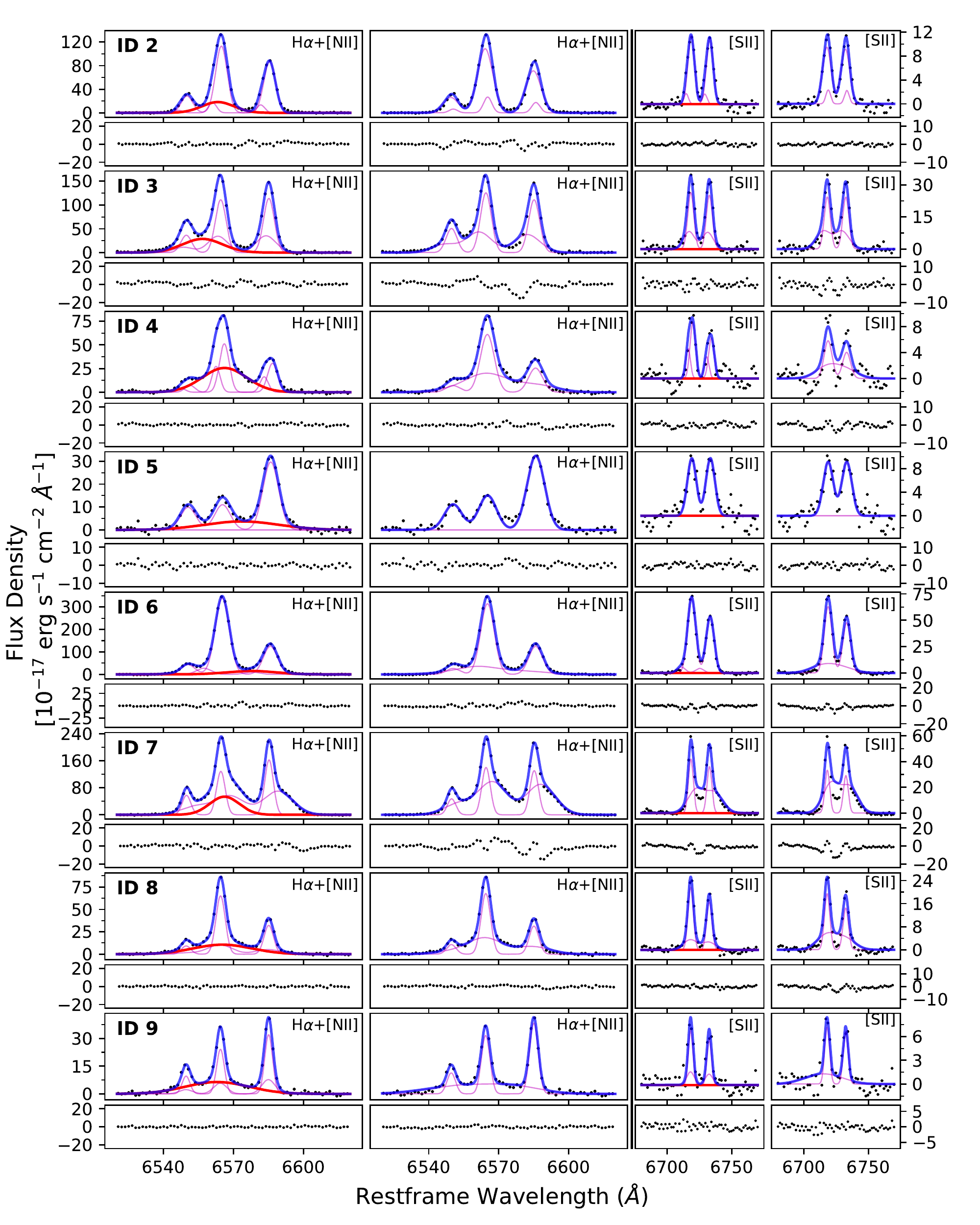}
    \caption{Spectral decomposition of the 24 broad H$\alpha$ candidate whose estimated M$_{\rm BH}$ is smaller than $10^{6}$ \msun.  The first and third column show the best fit model with H$\alpha$ broad component for each emission line region. The second and forth column show the best fit model without H$\alpha$ broad component. The stellar continuum subtracted spectrum is plotted black, and the blue line shows the best fit model composed of narrow Gaussian components (magenta) and H$\alpha$ broad component (red). At the bottom of each panel, the residuals between the spectrum and the best fit model are plotted in black. \label{fig:fitting_1_8}}
    \vspace{5mm}
\end{figure*}
%%%%%%%%%%%%%%%%%%%%%%%%%%%%%%%%%%%%%%%%%%%%%%%%%%%%%%%%%%%%%%%%%%%%%%%%%%%%%%%
%%%%%%%%%%%%%%%%%%%%%%%%%%%%%%%%%%%%%%%%%%%%%%%%%%%%%%%%%%%%%%%%%%%%%%%%%%%%%%%
%%% FIGURE %%%%%%%%%%%%%%%%%%%%%%%%%%%%%%%%%%%%%%%%%%%%%%%%%%%%%%%%%%%%%%%%%%%%
\begin{figure*}
    \centering
    \addtocounter{figure}{-1}
    \includegraphics[angle=0,width=175mm]{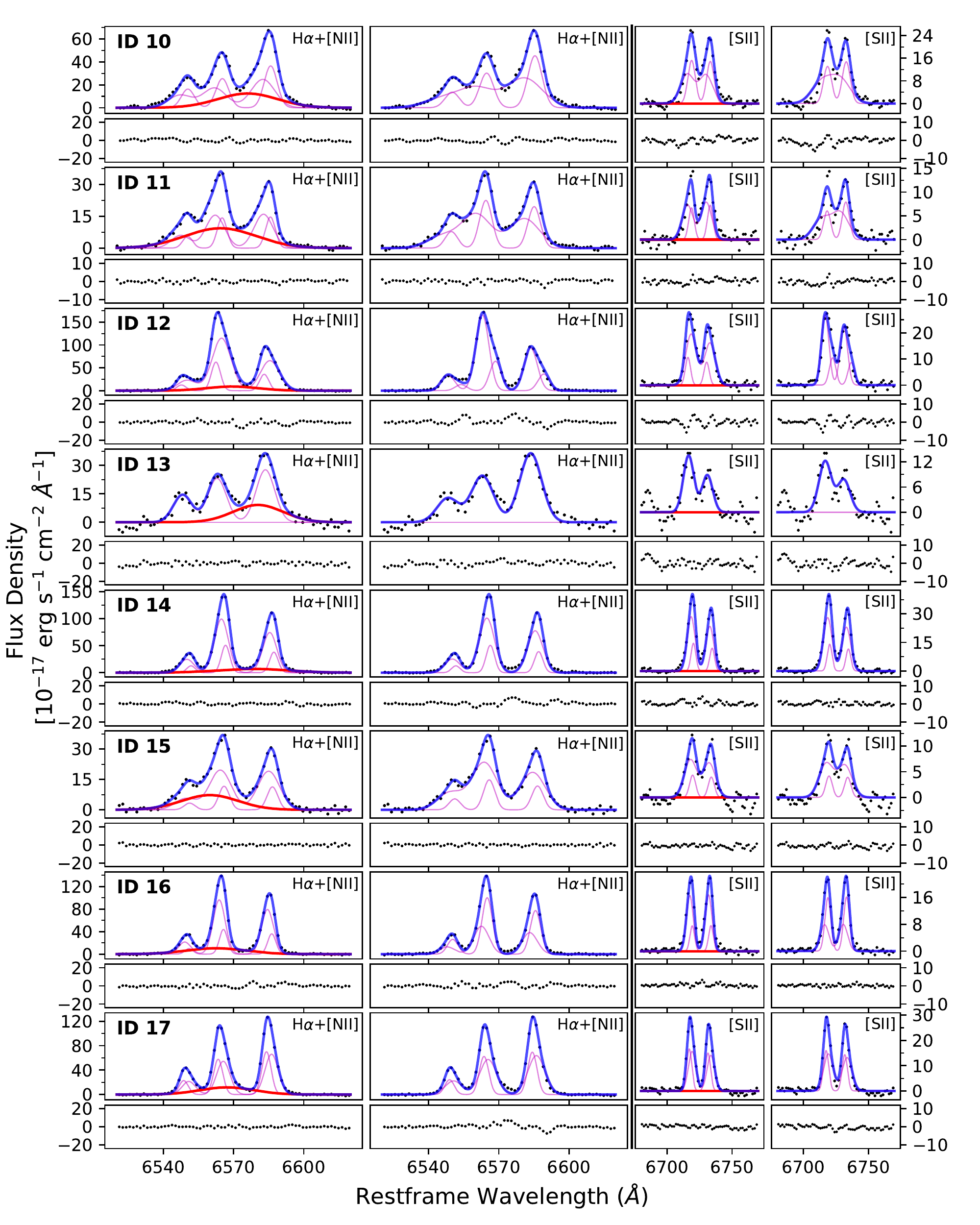}
    \caption{Continued \label{fig:fitting_9_16}}
    \vspace{5mm} %% add extra space ONLY when figures/tables are "colliding"!
\end{figure*}
%%%%%%%%%%%%%%%%%%%%%%%%%%%%%%%%%%%%%%%%%%%%%%%%%%%%%%%%%%%%%%%%%%%%%%%%%%%%%%%
%%%%%%%%%%%%%%%%%%%%%%%%%%%%%%%%%%%%%%%%%%%%%%%%%%%%%%%%%%%%%%%%%%%%%%%%%%%%%%%
%%% FIGURE %%%%%%%%%%%%%%%%%%%%%%%%%%%%%%%%%%%%%%%%%%%%%%%%%%%%%%%%%%%%%%%%%%%%
\begin{figure*}
    \centering
    \addtocounter{figure}{-1}
    \includegraphics[angle=0,width=175mm]{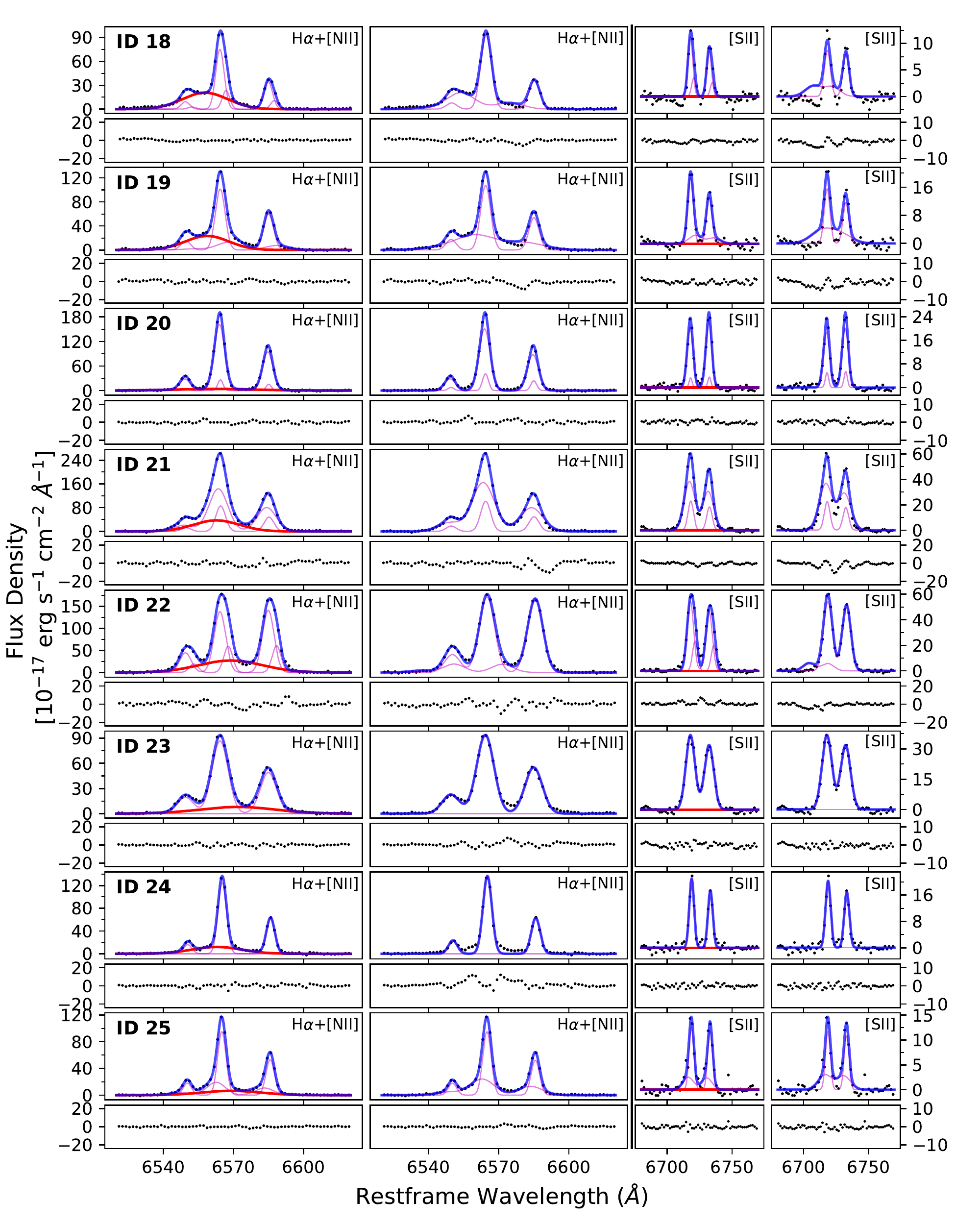}
    \caption{Continued \label{fig:fitting_17_24}}
    \vspace{5mm} 
\end{figure*}
%%%%%%%%%%%%%%%%%%%%%%%%%%%%%%%%%%%%%%%%%%%%%%%%%%%%%%%%%%%%%%%%%%%%%%%%%%%%%%%

\end{document}